\documentclass[3p,10pt]{elsarticle}

\journal{Computational and Mathematical Methods in Medicine}

\usepackage{epstopdf}
\usepackage{float}
\usepackage{amssymb}
\usepackage{graphicx}
\usepackage{amsmath}
\usepackage{mathptmx}
\usepackage{color} 
\usepackage{setspace} 
\usepackage{amsthm}
\usepackage{subfigure}
\usepackage{url}
\usepackage{array,multirow}
\usepackage{booktabs}
\usepackage{colortbl}
\usepackage{lscape}
\usepackage[
singlelinecheck=false 
]{caption}

\newcommand{\ffrac}[2]{\ensuremath{\frac{\displaystyle #1}{\displaystyle #2}}}
\newcommand{\p}{\partial}

\begin{document}

\begin{frontmatter}
\title{Parameter identifiability of a respiratory mechanics model in a preterm infant}

\author[label1]{Laura Ellwein Fix}

\address[label1]{Deptartment of Mathematics and Applied Mathematics\\
 Virginia Commonwealth University, Richmond, VA, USA\\
Email: lellwein@vcu.edu}

\begin{abstract}
The complexity of mathematical models describing respiratory mechanics has grown in recent years to integrate with cardiovascular models and incorporate nonlinear dynamics. However, additional model complexity has rarely been studied in the context of patient-specific observable data. This study investigates parameter identification of a previously developed nonlinear respiratory mechanics model (Ellwein Fix, PLoS ONE 2018) tuned to the physiology of 1 kg preterm infant, using local deterministic sensitivity analysis, subset selection, and gradient-based optimization. The model consists of 4 differential state equations with 31 parameters to predict airflow and dynamic pulmonary volumes and pressures generated under six simulation conditions.  The relative sensitivity solutions of the model state equations with respect to each of the parameters were calculated with finite differences and a sensitivity ranking was created for each parameter and simulation. Subset selection identified a set of independent parameters that could be estimated for all six simulations. The combination of these analyses produced a subset of 6 independent sensitive parameters that could be estimated given idealized clinical data. All optimizations performed using pseudo-data with perturbed nominal parameters converged within 40 iterations and estimated parameters within $\sim$8\% of nominal values on average. This analysis indicates the feasibility of performing parameter estimation on real patient-specific data set described by a nonlinear respiratory mechanics model for studying dynamics in preterm infants.\\
\end{abstract}


%

\end{frontmatter}


\section{Introduction}\label{intro}

Respiratory mechanics have been investigated mathematically for several decades using differential equations models that typically predict air pressure and flow in and between compartments representing aggregate features of the respiratory system. Models have grown in complexity from early compartmental models of dynamic volumes and pressures in the airways, lungs, chest wall, and intrapleural space~\cite{Golden73}. Successive models have built upon this foundation by including nonlinear resistances and compliances, viscoelastic components, and pulmonary circulation~\citep{Verbraak91,Liu98,Athan00}, and more recently been adapted to newborn animal physiology~\cite{LeRolle13}. We previously developed a dynamic nonlinear computational model of infant respiratory mechanics parameterized for the extremely preterm human infant~\citep{Ellwein18} to propose a mechanism of delayed progressive lung volume loss attributed to high chest wall compliance (floppiness)~\citep{Love53,Beltrand08,Kovacs15}.  Our model is the first known attempt to represent these dynamics in premature infants, and also depict the mitigating effects of expiratory laryngeal braking (grunting) and continuous positive airway pressure (CPAP) under simulated high and low chest wall compliance conditions. However, the parameter space contributing breathing dynamics, progressive volume loss, and the responses to interventions has not been explored. Given that ventilation assistance continues to fail in this population~\citep{Manley13,Bhandari13,Siew15} and with unknown etiology, this remains an area of continued study. 

Forward model simulations using parameter values obtained from experiments or population-based averages may provide insight into overall dynamics of a group, but estimating patient-specific parameters requires an optimization algorithm to find parameters that generate model output that best fits experimental data. In large and highly nonlinear physiological system with parameters numbering in the tens or hundreds and a scarcity of data describing relevant states, the optimization problem comes with two inherent challenges for obtaining unique parameter values. The first is parameter {\it sensitivity}, the impact of variation in parameter values on associated model output. A sensitivity analysis can examine the small local changes around each nominal parameter value or the global variability throughout the admissible parameter space~\citep{Eslami94,Karnavas93,LeRolle13,Sher13,Olsen18,Roosa19}. Parameter values may be only valid in a local region, or the behavior of a nonlinear component maybe be quasi-linear under a particular set of dynamic conditions. The second challenge is parameter {\it identifiability}, either due to structure of the model (structural) or the availability of data (practical)~\citep{Sher13,Olsen18,Kao18,Roosa19}. For example, two parameters that co-vary or depend on each other may not be able to have unique values estimated via an optimization algorithm. These questions are especially critical in the context of typical clinical data ``tracings'', which for assessment of respiratory mechanics may include only volumetric airflow as measured by a pneumotachograph, and pleural pressure as estimated by a pressure transducer in the esophagus. Data acquired under different experimental conditions makes it possible that a tracing from one of the two outputs may change significantly but the other output may show negligible differences, or tracings may be similar between conditions but mask different underlying dynamics. It is therefore critical to investigate which parameters most influence the model outputs under which conditions, and if any parameter dependencies exist that may allow for simplifying model components. 

Parameter sensitivity analysis and estimation have been used frequently in previous physiological modeling efforts, with the question of identifiability being explored more recently. Parameter estimation was performed in several linear respiratory mechanics models~\citep{Lutchen86,Verbraak91,Avanz97,Saatci08} but these were all linear and did not pose the challenges of nonlinear system. In the more sophisticated model of breathing in newborn lambs of Le Rolle et al~\cite{LeRolle13}, the parameter space was explored with an elementary effects algorithm which produced eight parameters identified with an evolutionary algorithm. Olsen et al examined several sensitivity analysis and parameter identification techniques in the context of increasingly complex biological models, finding that in an optimization of a cardiovascular model of blood flow and pressure, local sensitivity methods were preferable to global methods given a reasonable initial parameter set~\cite{Olsen18}. Additional studies on cardiovascular models include a local sensitivity analysis to reduce the size and parameter space of a compartmental model~\citep{Ellwein08}, a comparison of healthy and elderly groups to determine differential impact of parameters~\citep{Pope09}, and a cardiorespiratory model~\cite{Ellwein13}. Ipsen et al~\citep{Ipsen11} developed a subset selection technique based on singular value decomposition and QR factorization for parameter identifiability, motivated by the natural rank-deficient nature of these cardiovascular models; a related SVD-based technique was proposed by Sher et al~\cite{Sher13}. Raue et al~\cite{Raue13} systematically compared the performance of several optimization algorithms in the context of modeling of cellular dynamics, finding that a deterministic optimization algorithm was most efficient for parameterizing a differential equations assuming good initial parameter values. Derivative-free methods such as the simplex method Nelder-Mead are available for optimization, but generally are passed by in favor of gradient-based algorithms when a system of well-behaved differential equations system is being analyzed~\cite{Kelley99}.

Given the differential equation structure of the model studied here plus physiological knowledge of parameter initial guess values, local sensitivity analysis is applied together with the subset selection techniques developed by Ipsen et al to determine an independent sensitive parameter set over all simulation conditions. The novelty in this study comes from the application to multiple simulation conditions of dynamic breathing that have the majority of parameter values in common but several whose values vary between conditions. This suggests that a subset may be able to distinguish between two datasets acquired under different conditions, and in future investigations lead to uncovering physiological mechanisms leading to the observed dynamics. Therefore, the objective of this study is to find an independent sensitive parameter subset common to all simulations that may elucidate differences between six simulation scenarios describing the effects of grunting, CPAP, and no intervention during high and low chest wall compliance conditions. We begin with a brief description of the mathematical model and experimental setup. Then sensitivity analysis, subset selection, and special considerations important for analysis of breathing dynamics are described and an independent sensitive subset of candidate parameters is obtained for all simulations. Finally, we test the subset in a series of gradient-based optimizations to indicate the feasibility of parameter estimation using pseudo-data generated from the simulated outputs and perturbed nominal parameter sets. 


\begin{table}[!ht]
\caption{Descriptions of state variables and parameters.}
{\footnotesize
 \begin{tabular}{ll} 
\hline\noalign{\smallskip}
	Parameter/State 								& Physiologic description							\\ 
\noalign{\smallskip}\hline\noalign{\smallskip}
TLC [ml]									& Total lung capacity 																\\	
RV [ml]										& Residual volume  																	\\	
FRC [ml] 									& Functional residual capacity 										\\	
VC[ml]										& Vital capacity 																	\\	
RR [br/min] 							& Respiratory rate																\\	
$f$ [br/s]								& Respiratory frequency																\\	
$T$ [s]										& Duration of respiratory cycle												\\	
$V_T$ [ml]								& Tidal volume																	 \\		
$\dot{V}_E$ [ml/min]			& Minute ventilation															\\	
$\dot{V}_A$ [ml/s]				& Airflow																					\\	
$A_{mus}$ [cm H$_2$O]			& Muscle pressure amplitude												\\	
$P_{tm}$ [cm H$_2$O]				& Transmural pressure															\\	
$P_{A}$ [cm H$_2$O]				& Alveolar pressure																\\	
$P_{el}$ [cm H$_2$O]			& Lung elastic recoil (transpulmonary pressure)		\\	
$P_{ve}$ [cm H$_2$O]			& Viscoelastic component of pressure									\\
$P_{l,dyn}$ [cm H$_2$O]		& Dynamic pulmonary pressure															\\
$P_{pl}$ [cm H$_2$O]			& Pleural pressure																\\	
$P_{cw}$ [cm H$_2$O]			& Chest wall elastic recoil																\\	
$P_{mus}$ [cm H$_2$O]			& Respiratory muscle pressure												\\	
$C_A$ [ml/cm H$_2$O]			& Lung compliance																	\\	
$C_w$ [ml/cm H$_2$O]			& Chest wall compliance														\\	
$C_{rs}$ [ml/cm H$_2$O]		& Respiratory system compliance										\\	
$R_{rs}$ [cm H$_2$O s/L] 	& Respiratory system resistance										\\	
$\nu$											& Fraction of VC for chest wall relaxation volume 		\\	
$V_0$ [ml] 								& Chest wall relaxation volume												\\	
$\beta$										&	Baseline fraction of lung recruited at $P_{el}=0$	 \\
$\gamma$									&	Maximum fractional recruitment of lung							 \\
$\alpha$									&	Lower asymptote, fraction recruitment							  \\
$k$ [1/cm H$_2$O]					&	Lung elasticity coefficient												 \\
$c_F$	[cm H$_2$O]					&	Pressure at maximum lung recruitment							  \\
$d_F$	[cm H$_2$O]					&	Characterizes slope at maximum lung recruitment		 \\					
$c_w$ [cm H$_2$O]					& Transition point, chest wall compliance							\\	
$d_w$ [cm H$_2$O]					& Chest wall compliance slope coefficient		\\	
$c_c$ [cm H$_2$O]					& Pressure at peak collapsible airway compliance			\\	
$d_c$ [cm H$_2$O]					& Collapsible airway compliance slope coefficient	\\
$K_c$ [cm H$_2$O s/L]			& Collapsible airway resistance coefficient					\\	
$V_{c,max}$ [ml]					& Peak collapsible airway volume											\\	
$R_{s,m}$ [cm H$_2$O s/L]	& Minimum small airway resistance											\\	
$R_{s,d}$ [cm H$_2$O s/L]	& Change in small airway resistance										\\	
$K_s$ 										& Small airway resistance low pressure coefficient		\\	
$I_u$ [cm H$_2$O s$^2$/L]	& Upper airway inertance														\\		
$R_{u,m}$ [cm H$_2$O s/L]	& Laminar value, upper airway resistance						\\	
$K_u$ [cm H$_2$O s/L]			& Turbulent coefficient, upper airway resistance		\\	
$R_{u,mult}$							& Level of expiratory resistance increase 				\\
$C_{ve}$ [L / cm H$_2$O] 	& Lung viscoelastic compliance														\\
$R_{ve}$ [cm H$_2$O s/L] 	& Lung viscoelastic resistance												\\	
\noalign{\smallskip}\hline
\end{tabular}}
\label{tab:glossary}
\end{table}


\section{Methods}\label{sec:methods}

\begin{figure*}
\centering
\includegraphics[scale=0.85]{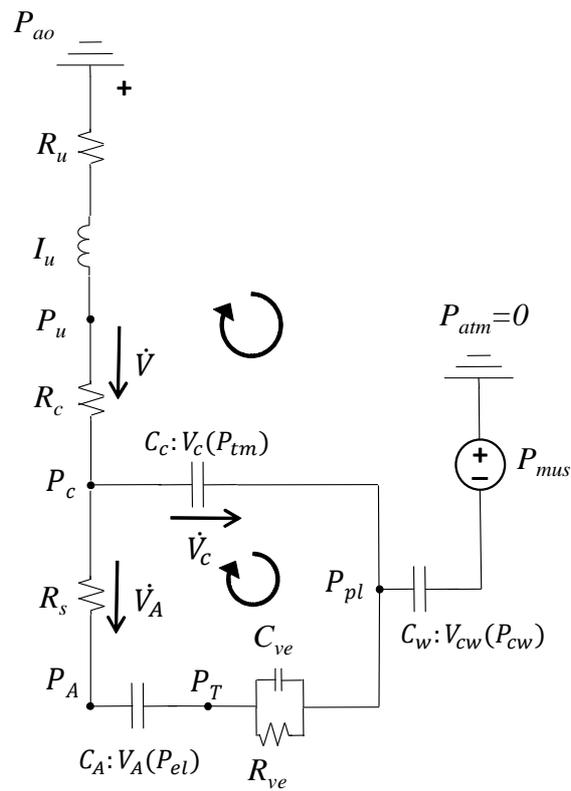}
\caption{Lumped-parameter respiratory mechanics model shown as its electrical analog, adapted from Ellwein Fix et al~\cite{Ellwein18}. Each compliant compartment $C$ has an associated volume $V$ as a function of the transmural pressures P across the compartment boundaries. Air flows $\dot{V}$ across resistances $R$ and inertance $I$ are positive in the direction of the arrows. Circular arrows indicate direction of loop summations used to derive the system of differential equations (Eq.~\ref{eq:states}). Subscripts: airway opening $ao$, upper $u$, collapsible $c$, small peripheral $s$, alveolar $A$, viscoelastic $ve$, lung elastic $el$, tissue $T$, transmural $tm$, pleural $pl$, chest wall $cw$, muscle $mus$.} 
\label{fig:model}
\end{figure*}

The compartmental respiratory mechanics model analyzed in this study here  was previously developed and parameterized specifically to investigate the dynamics of breathing in the extremely preterm infant weighing $\sim$1 kg, see Fig.~\ref{fig:model}. The naturally very high chest wall compliance of these infants has been hypothesized to be a major factor in clinically observed respiratory distress occurring in otherwise stable infants, thus the focus of the previous model was analyzing the differential impact of high vs. low chest wall compliance ($C_w$) on model state outputs under no intervention and the two simulated interventions of grunting (intrinsic) and CPAP (externally applied). We briefly describe the model here but also refer the reader to~\cite{Ellwein18} for full details.

\subsection{State equations}

The model describes dynamic volumes and pressures in the airways, lungs, chest wall, and intrapleural space between lungs and chest. A sigmoidal signal that represents a combined respiratory muscle pressure generated during spontaneous breathing drives the model. The model is designed using the volume-pressure analog of an electrical circuit, with states in terms of pressure $P(t)$ [cm H$_2$O] and volume $V(t)$ [ml] in and between air compartments and with volumetric flow rate and rate of change of compartmental volume represented as $\dot{V}(t)$ [ml/s] and $\frac{dV}{dt}$ respectively. Air pressure $P_i$ within volume $i$ is defined relative to the external atmospheric pressure, set as $P_{ext}=0$. 

The pressure $P_{ij}=P_i-P_j$ refers to the transmural pressure across a compliant boundary separating volumes $i$ and $j$.  Pressures $P_{ij}$ include transmural pressure between the compliant airways and the pleural space $P_{tm}=P_c-P_{pl}$, lung elastic recoil $P_{el}=P_A-P_T$, lung viscoelastic component $P_{ve}=P_T-P_{pl}$, and chest wall elastic recoil $P_{cw}=P_{pl}-P_{mus}$. Nonlinear compliance $C_i$ [ml/cm H$_2$O] of a compartment is described by $dV_i/dt=C_i(dP_{ij}/dt)$, resistance $R_i$ [cm H$_2$O$\cdot{}$s/ml] in the airways or tissues by $\dot{V}_i=(P_{i-1}-P_i)/R_i$, and inertial effects $I$ in the upper rigid airway (trachea) as $P_{i-1}-P_i=I\ddot{V}$. Conservation laws require that $V=V_{cw}=V_A+V_c$, in other words the total system volume equals the chest wall volume, which is the sum of the alveolar and compressible airway volumes. The resulting system of differential equations after summing pressures over loops and incorporating compliances and resistances is given by:
\begin{eqnarray}
\ddot{V}&:&\frac{d\dot{V}}{dt}=\frac{1}{I_u}\left(P_{ao}-P_u-R_u\dot{V}\right) \label{eq:states}\\
\dot{V}_c&:&\frac{dV_c}{dt}=\dot{V}-\dot{V}_A \nonumber\\
\dot{P}_{el}&:&\frac{dP_{el}}{dt}=\frac{\dot{V}_A}{C_A} \nonumber\\
\dot{P}_{ve}&:&\frac{dP_{ve}}{dt}=\frac{\dot{V}_A-(P_{ve}/R_{ve})}{C_{ve}} \nonumber
\end{eqnarray}

Variable and parameter descriptions are given in Table~\ref{tab:glossary}. All previously developed formulations describing the nonlinear compliances and resistances are summarized in Table~\ref{tab:functions}. The quantity $C_A$ is implicitly described by $V_A(P_{el})$ and was calculated exactly using symbolic computation as $\p V/\p P$. Breath-to-breath values for dynamic lung compliance $C_L$ and chest wall compliance $C_w$ are estimated during tidal breathing as $\Delta V/\Delta P$ from the curves $V_A(P_{el})$ and $V_{cw}(P_{cw})$ respectively, and compared with literature to lend support to simulated outputs (see Table~\ref{tab:vary}). Pressure $P_{mus}$ describes the effective action of the respiratory muscles that drives the model dynamics, with $P_{mus}$ negative in the outward direction. 

\begingroup
\renewcommand{\arraystretch}{1.7} 
\begin{table}[!ht]
\caption{Model functions for constitutive relations. See Ellwein Fix et al~\cite{Ellwein18} for detailed descriptions.}
{\small
\begin{tabular}{ll}	
\hline\noalign{\smallskip}
Description														& Function 																															\\ 
\noalign{\smallskip}\hline\noalign{\smallskip}
Upper airways resistance							& $R_u=R_{u,m}+K_u|\dot{V}|$ 																													\\
Collapsible airways resistance				& $R_c=K_c\left(\ffrac{V_{c,max}}{V_c}\right)^2$			 																\\
Small (peripheral) airways resistance	& $R_s=R_{s,d}\cdot e^{K_s(V_A-RV)/(TLC-RV)}+R_{s,m}$,\ \ \ \ $K_s<0$									\\ 
Collapsible airways volume compliance	& $V_c=\ffrac{V_{c,max}}{1+e^{-(P_{tm}-c_c)/d_c}}$  																	\\
Chest wall volume compliance					& $V_{cw}=RV+(V_0-RV) \ln \left(1+e^{P_{cw}/d_w}\right)/(\ln 2)$										\\
Lung tissue volume compliance					&	$V_A=V_{el}(P_{el})\cdot F_{rec}(P_{el})+RV$,\ \ \ \ where													\\
																			& \ \ \ \ \ \ \ \ \ \ $V_{el}=VC\cdot(1-e^{(-kP_{el})})$\ \ \ \ \ and											\\
																			& \ \ \ \ \ \ \ \ \ \ $F_{rec}=\alpha+\ffrac{\gamma-\alpha}{1+e^{-(P_{el}-c_F)/d_F}}$			\\
Lung viscoelastic recoil							& $C_{ve}\ffrac{dP_{ve}}{dt}=\dot{V}_A-P_{ve}/R_{ve}$																		\\
Diaphragm muscle driving pressure			& $P_{mus}=A_{mus}\cos (2\pi f t)-A_{mus}$																				\\
\noalign{\smallskip}\hline
\end{tabular}}
\label{tab:functions}
\end{table}
\endgroup

\begin{table}[!h]
\caption{Model parameters set to the same nominal values across all simulations. See Table~\ref{tab:glossary} for parameter descriptions.}
{\footnotesize
 \begin{tabular}{llccc} 
\hline\noalign{\smallskip}
Parameter 		& Value	& Formula 								& References 				\\ 
\noalign{\smallskip}\hline\noalign{\smallskip}
TLC 					&  63		& \textemdash							& \cite{Smith76,Donn98}	\\	
RV 						&  23		& \textemdash							& \cite{Smith76}	\\	
VC						&  40		&  TLC-RV									& \cite{Smith76,Donn98}	\\	
RR  					&  60		& \textemdash							& \cite{Donn98}	\\	
$f$ 					& 	1		& RR/60										& \textemdash	\\	
$T$ 					&  1		& $1/f$										& \textemdash	\\	
$\nu	$				& 0.25	&	\textemdash							& \cite{Donn98,Goldsmith11}	\\	
$V_0$ 				&  35		& $\nu\cdot$VC+RV					& \textemdash	\\	
$\beta$				&	0.01	&	estimated								& \cite{Hamlington16}  \\
$\gamma$			&		1		&	estimated								& \cite{Hamlington16}  \\
$\alpha$			&	-0.76	&	$\frac{(1+e^{c_F/d_F})\beta-\gamma}{e^{c_F/d_F}}$	& \cite{Hamlington16}  \\
$k$ 					&	0.07	&	estimated								& \cite{Ferreira11,Hamlington16}  \\
$c_F$					&		0.1	&	estimated								& \cite{Hamlington16}  \\
$d_F$					&	0.4		&	estimated								& \cite{Hamlington16}  \\					
$c_w$ 				&  0		& estimated								& \textemdash	\\	
$c_c$ 				&  4.4	& estimated from adult		& \cite{Liu98}	\\	
$d_c$ 				&  4.4	& estimated from adult		& \cite{Liu98}		\\
$K_c$ 				&  0.1	& estimated from adult		& \cite{Athan00}	\\	
$V_{c,max}$ 	&  2.5	& estimated as dead space	& \cite{Donn98,Neumann15}	\\	
$R_{s,m}$ 		&  12		& \textemdash							& \cite{Ratjen92,Singh12}	\\	
$R_{s,d}$ 		&  20		& estimated from adult		& \cite{Athan00}	\\	
$K_s$ 				&  -15	& estimated from adult		& \cite{Athan00}	\\	
$I_u$ 				&  0.33	& \textemdash							& \cite{Singh12,LeRolle13}\\		
$C_{ve}$  		& 0.005 & estimated from adult 		& \cite{Athan00}				\\		
$R_{ve}$  		& 20		& estimated from adult		& \cite{Athan00}	\\	
$R_{u,m}$     & 20    & estimated								& \citep{Mortola87,Singh12} \\
$K_u$					& 60		& estimated								& \citep{Mortola87,Athan00,Singh12}	\\	
\noalign{\smallskip}\hline
\end{tabular}}
\label{tab:SSparams}
\end{table}


\subsection{Simulation conditions and model parameters}\label{sec:parameters}

Simulation conditions were chosen to demonstrate high and low chest wall compliance conditions without any intervention and with two simulated interventions, totaling six sets of conditions. The dynamics of five tidal (steady) breathing cycles prior to the onset of progressive volume loss were analyzed. Minute ventilation $\dot{V}_E$ was set at 360 ml/min~\citep{Donn98} for each simulation to enable comparison of dynamics under the same ventilation requirement. This was achieved by imposing a respiratory rate (RR) of 60 br/min and setting respiratory muscle pressure amplitude ($A_{mus}$) to obtain constant tidal volume ($V_T$) of 6 ml~\citep{Pandit00,Habib03,Schmalisch05}. Grunting, simulated as increased airway resistance during expiration, is implemented by multiplying the entire $R_u$ expression by $R_{u,mult}=10$ when $\dot{V}<0$. CPAP is applied by setting $P_{ao}=5$.

Nominal parameters for model equations in Table~\ref{tab:functions} were tuned in the previous study such that resulting lung and chest wall compliance curves produced functional residual capacity (FRC), states and dynamic compliances comparable to reported literature values. (See Ellwein Fix et al.~\cite{Ellwein18} for greater detail.) Functional residual capacity, the volume at 0 respiratory pressure, was calculated using a nonlinear solver as the volume where static $P_{cw}+P_{el}=0$, thus taking on different values based on the chest wall compliance curve. The majority of parameters are assigned the same nominal values across all simulations, as shown in Table~\ref{tab:SSparams}.  However, the four parameters given in Table~\ref{tab:vary} take on values that differ for each simulation to generate the requisite minute ventilation.  The single degree of freedom $d_w$ for the chest wall compliance curve differentiates between high and low chest wall compliance. Table~\ref{tab:vary} also reports dynamic lung and chest wall compliances as the approximate slopes of the parameterized $V_A(P_{el})$ and $V_{cw}(P_{cw})$ curves during tidal breathing for each simulation to show consistency with prior studies.

The system of differential equations~\eqref{eq:states}, together with the constitutive relations in Table~\ref{tab:functions}, were solved using MATLAB R2016b (MathWorks, Natick, MA) with the differential equations solver \verb|ode15s| using a tolerance of $1\mathrm{e}{-8}$. Initial conditions for $\dot{V}$, $P_{ve}$, and $V_c$ were set at 0, 0, and 0.0001 respectively ~\citep{Liu98}. Initial conditions for $P_{el}$ were 0.954 and 2.015 for high and low chest wall compliance, calculated as the pressure at which FRC was established.

The full parameter set $\mu$ is comprised of parameters that have the same nomimal values across all simulations plus four parameters with values that differ between simulations, given in Tables~\ref{tab:SSparams} and~\ref{tab:vary}. The full set $\mu$ is split into two groups such that $\mu=\mu_{fix}\cup\mu_{inc}0$. The parameters of the group
\begin{equation}
\mu_{fix}=\{TLC,RV,VC,FRC,RR,f,T,V_0,\alpha,\nu\,c_w\}
\end{equation}
were kept fixed and not included in the sensitivity analysis because they were estimated {\it a priori} for an idealized subject or redundancies were already known. The remaining group of 20 parameters that makes up $\mu_{inc}$ is analyzed with the techniques of Sections~\ref{sec:SA} and~\ref{sec:SS}. Since $\mu_{inc}$ contains $A_{mus},d_w,R_{u,mult}$ that take on different nominal values for each simulation, each sensitivity analysis is relative to its own local parameter space. As a consequence, output tracings that are similar but generated from different simulations may still may be sensitive to different parameters in the full parameter set based on the underlying nonlinear dynamics.

\begin{table}[!h]
\caption{Model parameters set to different nominal values between simulations based on a minute ventilation of $\dot{V}_E=360 ml/min$, including dynamic lung and chest wall compliances $C_L$ and $C_w$. See Table~\ref{tab:glossary} for parameter descriptions.} 
{\scriptsize
 \begin{tabular}{lccccccc} 
\hline\noalign{\smallskip}
\multirow{2}{*}{}					& \multicolumn{3}{c}{High $C_w$}								& \multicolumn{3}{c}{Low $C_w$} 								& \multirow{2}{*}{References} 				\\ 
\noalign {\smallskip}
													&  normal $R_u$ & increased $R_u$	& $P_{ao}=5$	& normal $R_u$ & increased $R_u$	& $P_{ao}=5$	& 									\\ 
\noalign{\smallskip}\hline\noalign{\smallskip}
{\it Model Parameters}		&								&									&							&								&									&							&																\\
FRC$^\text{a}$	 									&  24.9					& 24.9 						& 24.9				& 28.1					& 28.1						& 28.1				& \citep{Smith76,Donn98,Thomas04}	\\	
$A_{mus}$ 								&  1.85					& 3.20 						& 2.21				& 2.78 					& 3.80 						& 2.76				& \textemdash	\\	
$d_w$ 										&  0.48					& 0.48 						& 0.48				& 2.4 					& 2.4 						& 2.4					& \textemdash	\\	
$R_{u,mult}$							&  1						& 10 							& 1						& 1							& 10							& 1						& \textemdash	\\
{\it Simulated Outputs}		&								&									&							&								&									&							&																\\
$C_L$ 										& 2.7						& 2.1							&	1.7					& 2.3						&	2.1							&	1.8					& \citep{Gerhardt80,Mortola87,Pandit00}	\\	
$C_w$ 										& 9.9						&	16.0						&	20.4				&	2.7						&	3.3							& 3.9					& \citep{Gerhardt80,Mortola87}	\\	
\noalign{\smallskip}\hline
\end{tabular}
\begin{flushleft} $^\text{a}$FRC is calculated by solving $P_{el}|_{FRC}+P_{cw}|_{FRC}=0$. 
\end{flushleft}}
\label{tab:vary}
\end{table}


\subsection{Sensitivity analysis}\label{sec:SA}

Local sensitivity analysis is performed on the nominal, or true, parameter set $\mu^*$ with values specific to each simulation such that there are six local parameter spaces to be analyzed. Each parameter space has different nominal values for the four parameters given in Table~\ref{tab:vary}, but the same values for the remaining parameters. We used a differential equation analysis approach~\citep{Eslami94,NCSU} for calculating time-dependent sensitivities $\p y_i(t)/\p \mu_j$ for each simulation and parameter, calculated a single scalar sensitivity value for each, then ranked sensitivities for each simulation condition and overall. 

 The two model outputs generated an output vector of length $4N$:
\begin{equation}
y=[\dot{V}(t_1),...,\dot{V}(t_{2N}),P_{pl}(t_1),...,P_{pl}(t_{2N})]^T
\end{equation}
where $N$ is the number of time points in a single breathing cycle. The third and fourth full cycles were used to ensure that transient behavior was excluded. Note that $P_{pl}$ is not a state variable in the system of differential equations, but is obtained by $P_{pl}=P_{cw}(V_{cw})+P_{mus}$ where $V_{cw}=V_A+V_c$. To avoid problems when output values are near or at 0 which would produce infinite sensitivities (such as from using the airflow time series oscillating around 0)~\citep{Bahill80,Karnavas93,Wu10}, we scale each component of $y$ using its maximum value of the output in absolute value to obtain a scaled sensitivity matrix $\hat{S}$:
\begin{eqnarray}
\hat{y}&=&\left[\frac{\dot{V}(t_1)}{V_{max}},...,\frac{\dot{V}(t_{2N})}{V_{max}},\frac{P_{pl}(t_1)}{P_{pl,max}},...,\frac{P_{pl}(t_{2N})}{P_{pl,max}}\right]^T\\
  \left. \hat{S}(t,\mu)\right|_{\mu_{inc}=\mu^*} &= &
  \left. \frac{\p \hat{y}_i(t,\mu)}{\p \mu_j } \right|_{\mu=\mu^*}.
\end{eqnarray} 
 A non-dimensional relative sensitivity is then obtained by multiplying by the parameter value. Thus the relative sensitivity $S_{ij}$ of output $y_i$ to parameter $\mu_j$ at nominal parameter set $\mu^*$ is defined as
\begin{equation}
  \left. S_{ij}(t,\mu)\right|_{\mu_{inc}=\mu^*} = 
  \left.\mu_j \frac{\p \hat{y}_i(t,\mu)}{\p \mu_j} \right|_{\mu=\mu^*}.
\end{equation} 
Derivatives of $y$ with respect to $\mu_j$ were computed with a forward difference approximation using a difference increment of $\epsilon_j=1\mathrm{e}{-4}$~\cite{Ipsen11,NCSU}. 

To obtain a scalar value for ranking, we computed composite sensitivities $S_j$ using the standard 2-norm for each of the six simulations,
\begin{equation}
\left. S_{j}= \left\|S_{ij}(t,\mu)\right\|_2. \right.
\end{equation}
Composite sensitivities $S_j$ are tabulated in two ways. First, $S_j$ are ranked for each parameter and averaged across the six simulations to obtain an average ranking. Second, $S_j$ are first averaged across the set of six simulations to obtain a measure of sensitivity for each parameter, then graphically depicted in order of sensitivity. A point at which the sensitivities show a sharp gap or drop may be identified as the set of ``sensitive'' parameters, which will be referred to as $\mu_{sens}$.


\subsection{Subset selection}\label{sec:SS}

Discerning the relative impact of a parameter on output states does not identify any dependencies or redundancies between model parameters~\citep{Burth99,Heldt04,Pope09}. Identifying a set of independent parameters for the set of simulations may make future parameter estimation of patient-specific data sets more tractable and allow for reducing, linearizing, or simplifying model components. We use a subset selection method based on singular value decomposition and QR factorization~\citep{Ipsen11,NCSU} that addresses the question of practical identifiability, i.e. determining a set of independent parameters that are identifiable given limited experimental data. A brief description follows.

Subset selection begins with computing the singular value decomposition of the sensitivity matrix $\hat{S}|_{\mu_{inc}=\mu^*}=U\Sigma V^T$ where $\Sigma$ is a diagonal matrix of singular values in decreasing order and $V$ contains the corresponding right singular vectors. We predict a numerical rank $\rho$ which indicates the number of maximally independent columns of $S$, using a prescribed $\epsilon$ such that $\sigma_{\rho}/\sigma_1\ > \epsilon\geq 10\epsilon_J$, where $\sigma_1$ is the largest singular value and $\epsilon_J$ is the Jacobian finite difference approximation increment (thus giving the cutoff as $1\mathrm{e}{-3}$). The numerical rank is equivalent to the number of parameters that can be identified given the model output $y_i(\mu)$ and is used to partition $V=[V_{\rho}V_{n-\rho}]$ where $n$ is the total number of parameters analyzed. The particular parameters associated with the $\rho$ largest singular values are found using QR-decomposition with column pivoting. The permutation matrix $P$ that results from the decomposition $V_{\rho}^TP=QR$ is applied to reorder the parameter vector $\mu_{inc}$ to obtain $\tilde{\mu}^*=P^T\mu^*$, which is partitioned as $\tilde{\mu}^*=\tilde{\mu}^*_{\rho}\tilde{\mu}^*_{n-\rho}$. The vector $\tilde{\mu}^*_{\rho}$ then constitutes an independent set of model parameters that are estimable as part of a reduced-order optimization problem 
\begin{equation}
\arg \min \limits_{\mu} J(\tilde{\mu}^*_{\rho})
\end{equation}
where $J$ is the least-squares cost, while parameters $\tilde{\mu}^*_{n-\rho}$ remain fixed at baseline estimates. 

The subset selection method is applied to each of the six simulations. Given that each simulation condition is expected to produce a different independent subset based on differing underlying parameter values, the six subsets are tabulated to examine which parameters $\tilde{\mu}_{0,\rho}$ are chosen for most or all simulations to obtain $\mu_{sub}$. It is unlikely these directly overlap with the sensitive parameter subset, therefore $\mu_{sub}$ are compared against the set $\mu_{sens}$ found from sensitivity rankings to obtain an independent sensitive parameter subset $\mu_{est}$ suitable for optimization~\cite{Pope09}. 


\subsection{Optimization and parameter estimation}\label{sec:opt}
The objective of numerical optimization is to obtain an optimal set of parameter values that generates model output that best represents movel output. In the absence of clinical data, the feasibility of estimating optimal parameter values for the independent sensitive subset $\mu_{est}$ was demonstrated using pseudo-data and perturbed parameter values $\mu_0$ as initial guesses. Model outputs of airflow $\dot{V}$ and pleural pressure $P_{pl}$ were initially created by simulating the forward model with the nominal parameter sets for each of the six simulations. Three variations of pseudo-data were then generated by the addition of Gaussian noise at levels of $2\%$, $5\%$, and $10\%$. Perturbed parameters values $\mu_0$ for each optimization were generated by multiplying each parameter in $\mu_{est}$ by a value drawn from a uniform distribution between 0.5 and 1.5. Parameters not in $\mu_{est}$ that were not to be optimized were kept constant at nominal values. Optimizations were performed on 100 realizations of pseudo-data and perturbed parameter set combinations for each of the six simulations and three levels of noise. 

 Given the sets of simulated pseudo-data, the numerical optimization minimized the minimized the least-squares cost function $J$:
\begin{eqnarray}
  J = \sum_{i=1}^{2N} \left|\frac{\dot{V}^m(t_i:\mu)-\dot{V}^d_i}{\dot{V}_{max}^d}\right|^2 + \sum_{i=1}^{2N} \left|\frac{P_{pl:\mu}^m(t_i)-P_{pl,i}^d}{P_{pl,max}^d}\right|^2 
\label{eq:min_prob}
\end{eqnarray}
Superscripts {\it d,m} refer to the data and model respectively, and subscript {\it max} denotes the maximum in absolute value of each data set. Pseudo-data and perturbed nominal values of the independent sensitive parameter subset were input into a bound-constrained Levenberg-Marquardt (L-M) optimization algorithm with trust region~\citep{Ipsen11,NCSU,Kelley99}, with regularization parameter $\nu=0.2$. Lower and upper bounds for parameter constraints in L-M were set at $0.5\mu_0$ and $2\mu_0$ respectively. Each parameter value is scaled during the optimization by the difference of the bounds for that parameter. The L-M algorithm terminates with tolerance of $1\mathrm{e}{-4}$ based on gradient norm $||\nabla J(\mu)||$ or residual $J(\mu)$, i.e. the iteration continues until one has fallen below the tolerance for any of the convergence criteria. Mean and standard deviation of optimized parameters were reported and compared against true nominal parameter values $\mu_*$.


\section{Results}\label{sec:results}

\begin{figure}
\includegraphics[scale=0.45]{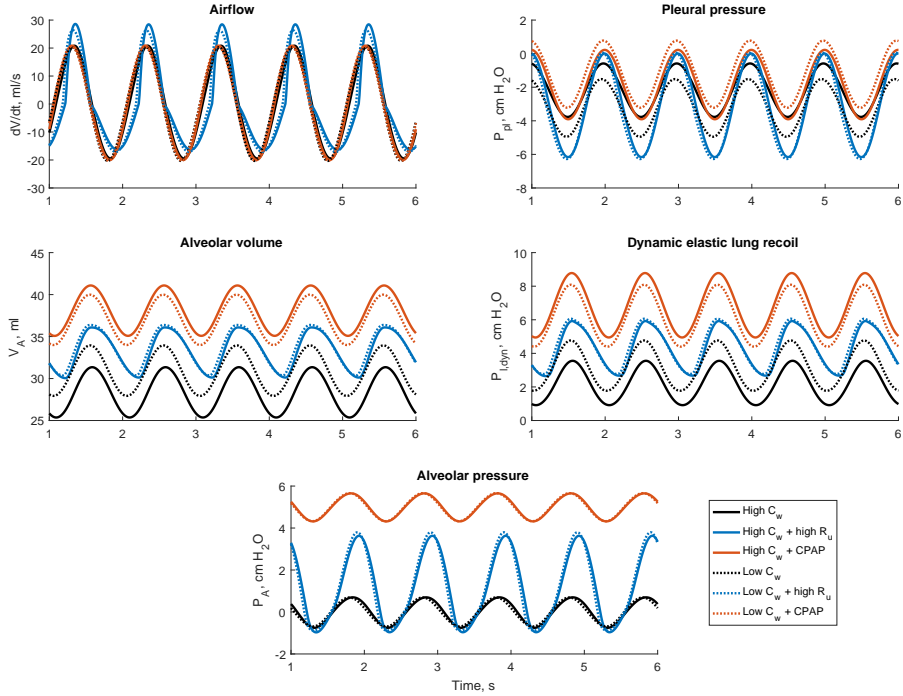}
\caption{Five dynamic period state variables shown in steady-state under six simulation conditions: high and low chest wall compliance, with and without CPAP and increased expired $R_u$. (For interpretation of colors in the legend, the reader is referred to the online version.)} 
\label{fig:states}
\end{figure}

Fig.~\ref{fig:states} gives the steady-state simulated tracings for $P_{pl}$, $\dot{V}$, $V_A$, $P_{l,dyn}$, and $P_A$ for all six simulations as described previously in Section~\ref{sec:parameters}. A typical clinical setup would only obtain data for $P_{pl}$, $\dot{V}$, and possibly a tracing for $V_A$ that resets to zero at each breathing cycle instead of actual end-expiratory lung volume (EELV). Note that $\dot{V}$ tracings are nearly identical under low and high $C_w$ conditions with and without CPAP, whereas with $R_u$ the tracings are slightly higher and shifted right $\sim$1 sec. Under no intervention (black lines), decreasing $C_w$ shifts the $P_{pl}$ curve down by about 1 cm H$_2$0, indicating a greater pleural pressure resulting from respiratory muscle activation and translating to higher alveolar volume. However, adding simulated CPAP to the high $C_w$ scenario (orange lines), the maximum negative $P_{pl}$ does not change considerably but the maximum $P_{pl}$ at end expiration increases by about 1 cm H$_2$0. Decreasing $C_w$ (dotted orange line) actually puts tidal breathing in a place above $V_0$ where $P_{pl}$ curve shifts up. 

The greatest effects from the simulated interventions can actually be seen in the tracings of $V_A$,  $P_{l,dyn}$, and $P_A$, though neither $P_{l,dyn}$ nor $P_A$ are data normally accessible clinically. The differences in $V_A$ under interventions are all vertical shifts reflecting different EELV, which would not be captured by clinical data tracings reset to zero volume at each breath. Tracings for $P_{l,dyn}$ qualitatively follow similar shifts as $V_A$. CPAP appears to greatly increase the tracing for $P_A$ where high $R_u$ stretches it, noting however that low chest wall compliance appears to have negligible effect on steady-state dynamics of $P_A$ vs high compliance. 

\subsection{Parameter identification}

\begin{figure}[h]
\includegraphics[scale=0.75]{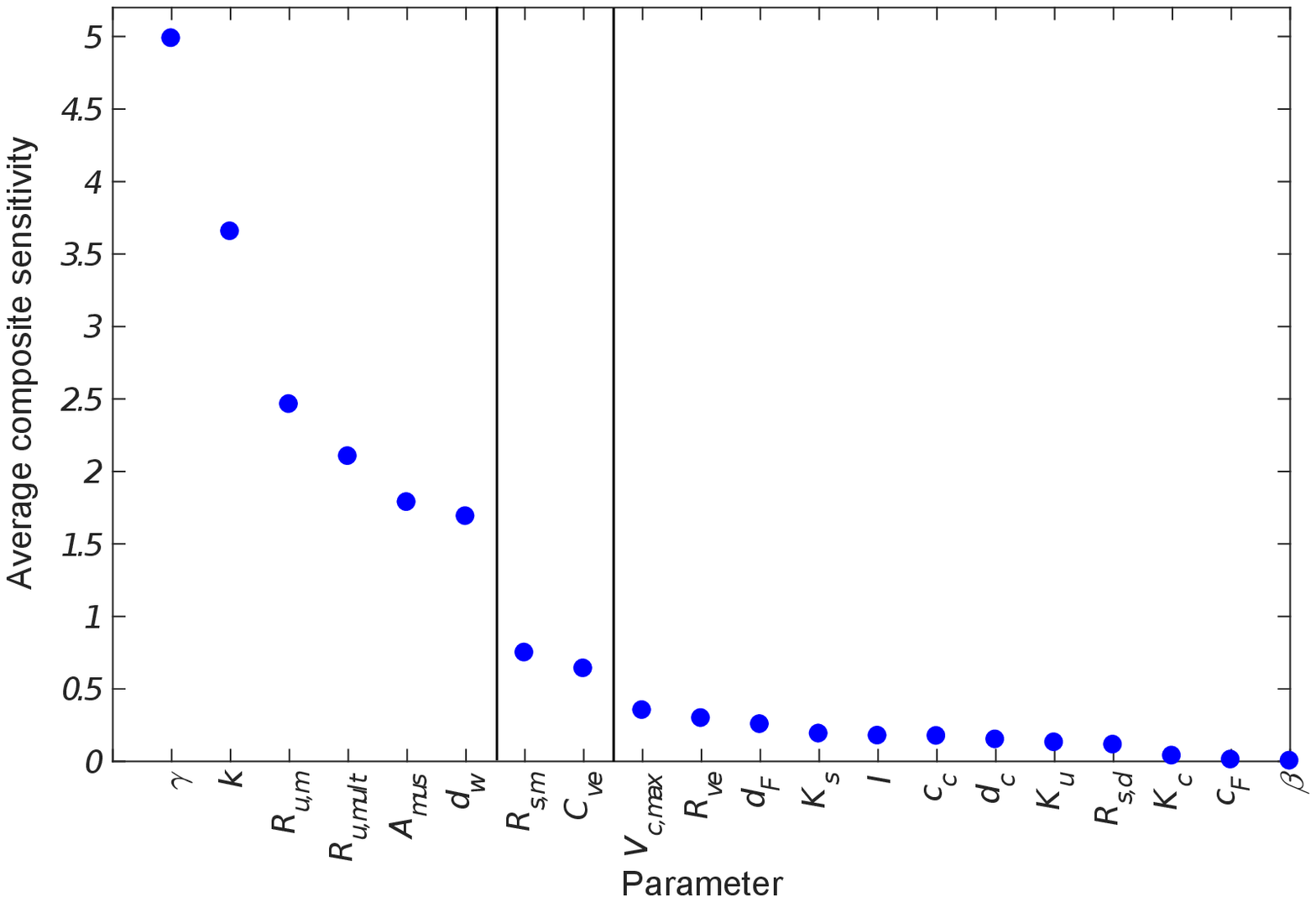}
\caption{Composite sensitivities $S_j$ averaged across the six simulations. Parameters are ordered by decreasing average sensitivity. Vertical lines indicate noticeable gaps in the ordering after the first two parameters, c.f. Table~\ref{tab:sens_rank}. } 
\label{fig:composite}
\end{figure}

Table~\ref{tab:sens_rank} gives the rankings for all six simulation conditions as described previously in Section~\ref{sec:SA}. For each individual simulation, 1 is the highest ranking and 20 is the lowest. Cells are also shaded for better visualizing the comparison of rankings based on simulation. Average rankings in the right-most column are reported as the mean across all six simulations. Fig.~\ref{fig:composite} shows actual average sensitivities across the six simulations displayed according to declining sensitivity. It is clear from both depictions that $\beta$, $c_F$, and $K_c$ rank consistently as low sensitivity parameters in all simulation cases and composite. Parameter $\gamma$ ranks overall as the most sensitive in all cases. Parameters $k$ and $R_{um}$ rank second and third overall respectively. The next two parameters $A_{mus}$ and $R_{u,mult}$ rank fourth and fifth by average ranking, but fifth and fourth by average sensitivity. A noticeable jump in average sensitivity occurs after the sixth parameter, giving the top six sensitive parameters as $\gamma$ (maximum fractional recruitment), $k$ (lung elasticity coefficient), $R_{u,m}$ (laminar upper airway resistance), $R_{u,mult}$ (level of airway braking), $A_{mus}$ (muscle pressure amplitude), and $d_w$ (chest wall compliance slope coefficient). A smaller jump occurs after the 8th parameter which would include $R_{s,m}$ (minimum small airway resistance) and $C_{ve}$ (viscoelastic compliance) as possible sensitive parameters.

\begin{table}[htbp]
  \centering
	\setlength{\tabcolsep}{0.3em}
  \caption{Sensitivity rankings for each of the six simulations, with an average ranking in the right-most column. Highlighting in cells corresponds with relative rankings. Horizontal lines indicate the first and second largest gaps in the rankings after the first two parameters, c.f. Fig.~\ref{fig:composite}.}
    {\footnotesize
		\begin{tabular}{l|ccc|ccc|c}
    \toprule
          & \multicolumn{3}{c|}{High $C_w$} & \multicolumn{3}{c|}{Low $C_w$} & Average \\
    \multicolumn{1}{c|}{Parameter} & normal $R_u$ & increased $R_u$ & $P_{ao}=5$ & normal $R_u$ & increased $R_u$ & $P_{ao}=5$ & ranking \\
    \midrule
    $\gamma$ & 1     & 1     & 1     & 1     & 1     & 1     & 1.0 \\
    $k$   & \cellcolor[rgb]{ .976,  .976,  .976}2 & \cellcolor[rgb]{ .922,  .922,  .922}4 & \cellcolor[rgb]{ .976,  .976,  .976}2 & \cellcolor[rgb]{ .949,  .949,  .949}3 & \cellcolor[rgb]{ .922,  .922,  .922}4 & \cellcolor[rgb]{ .976,  .976,  .976}2 & \cellcolor[rgb]{ .953,  .953,  .953}2.8 \\
    $R_{u,m}$ & \cellcolor[rgb]{ .922,  .922,  .922}4 & \cellcolor[rgb]{ .976,  .976,  .976}2 & \cellcolor[rgb]{ .949,  .949,  .949}3 & \cellcolor[rgb]{ .898,  .898,  .898}5 & \cellcolor[rgb]{ .976,  .976,  .976}2 & \cellcolor[rgb]{ .898,  .898,  .898}5 & \cellcolor[rgb]{ .937,  .937,  .937}3.5 \\
    $A_{mus}$ & \cellcolor[rgb]{ .949,  .949,  .949}3 & \cellcolor[rgb]{ .898,  .898,  .898}5 & \cellcolor[rgb]{ .871,  .871,  .871}6 & \cellcolor[rgb]{ .976,  .976,  .976}2 & \cellcolor[rgb]{ .871,  .871,  .871}6 & \cellcolor[rgb]{ .922,  .922,  .922}4 & \cellcolor[rgb]{ .914,  .914,  .914}4.3 \\
    $R_{u,mult}$ & \cellcolor[rgb]{ .871,  .871,  .871}6 & \cellcolor[rgb]{ .949,  .949,  .949}3 & \cellcolor[rgb]{ .922,  .922,  .922}4 & \cellcolor[rgb]{ .871,  .871,  .871}6 & \cellcolor[rgb]{ .949,  .949,  .949}3 & \cellcolor[rgb]{ .871,  .871,  .871}6 & \cellcolor[rgb]{ .906,  .906,  .906}4.7 \\
    $d_w$ & \cellcolor[rgb]{ .898,  .898,  .898}5 & \cellcolor[rgb]{ .871,  .871,  .871}6 & \cellcolor[rgb]{ .82,  .82,  .82}8 & \cellcolor[rgb]{ .922,  .922,  .922}4 & \cellcolor[rgb]{ .898,  .898,  .898}5 & \cellcolor[rgb]{ .949,  .949,  .949}3 & \cellcolor[rgb]{ .894,  .894,  .894}5.2 \\
    \midrule
    $R_{s,m}$ & \cellcolor[rgb]{ .843,  .843,  .843}7 & \cellcolor[rgb]{ .843,  .843,  .843}7 & \cellcolor[rgb]{ .898,  .898,  .898}5 & \cellcolor[rgb]{ .843,  .843,  .843}7 & \cellcolor[rgb]{ .843,  .843,  .843}7 & \cellcolor[rgb]{ .82,  .82,  .82}8 & \cellcolor[rgb]{ .851,  .851,  .851}6.8 \\
    $C_{ve}$ & \cellcolor[rgb]{ .82,  .82,  .82}8 & \cellcolor[rgb]{ .82,  .82,  .82}8 & \cellcolor[rgb]{ .843,  .843,  .843}7 & \cellcolor[rgb]{ .82,  .82,  .82}8 & \cellcolor[rgb]{ .82,  .82,  .82}8 & \cellcolor[rgb]{ .792,  .792,  .792}9 & \cellcolor[rgb]{ .82,  .82,  .82}8.0 \\
        \midrule
		$V_{c,max}$ & \cellcolor[rgb]{ .686,  .686,  .686}13 & \cellcolor[rgb]{ .741,  .741,  .741}11 & \cellcolor[rgb]{ .792,  .792,  .792}9 & \cellcolor[rgb]{ .714,  .714,  .714}12 & \cellcolor[rgb]{ .792,  .792,  .792}9 & \cellcolor[rgb]{ .843,  .843,  .843}7 & \cellcolor[rgb]{ .761,  .761,  .761}10.2 \\
    $R_{ve}$ & \cellcolor[rgb]{ .714,  .714,  .714}12 & \cellcolor[rgb]{ .792,  .792,  .792}9 & \cellcolor[rgb]{ .765,  .765,  .765}10 & \cellcolor[rgb]{ .765,  .765,  .765}10 & \cellcolor[rgb]{ .765,  .765,  .765}10 & \cellcolor[rgb]{ .741,  .741,  .741}11 & \cellcolor[rgb]{ .757,  .757,  .757}10.3 \\
    $d_F$ & \cellcolor[rgb]{ .792,  .792,  .792}9 & \cellcolor[rgb]{ .714,  .714,  .714}12 & \cellcolor[rgb]{ .557,  .557,  .557}18 & \cellcolor[rgb]{ .792,  .792,  .792}9 & \cellcolor[rgb]{ .663,  .663,  .663}14 & \cellcolor[rgb]{ .584,  .584,  .584}17 & \cellcolor[rgb]{ .682,  .682,  .682}13.2 \\
    $I$   & \cellcolor[rgb]{ .635,  .635,  .635}15 & \cellcolor[rgb]{ .765,  .765,  .765}10 & \cellcolor[rgb]{ .686,  .686,  .686}13 & \cellcolor[rgb]{ .584,  .584,  .584}17 & \cellcolor[rgb]{ .741,  .741,  .741}11 & \cellcolor[rgb]{ .663,  .663,  .663}14 & \cellcolor[rgb]{ .678,  .678,  .678}13.3 \\
    $c_c$ & \cellcolor[rgb]{ .608,  .608,  .608}16 & \cellcolor[rgb]{ .608,  .608,  .608}16 & \cellcolor[rgb]{ .741,  .741,  .741}11 & \cellcolor[rgb]{ .663,  .663,  .663}14 & \cellcolor[rgb]{ .686,  .686,  .686}13 & \cellcolor[rgb]{ .765,  .765,  .765}10 & \cellcolor[rgb]{ .678,  .678,  .678}13.3 \\
    $K_s$ & \cellcolor[rgb]{ .765,  .765,  .765}10 & \cellcolor[rgb]{ .635,  .635,  .635}15 & \cellcolor[rgb]{ .635,  .635,  .635}15 & \cellcolor[rgb]{ .741,  .741,  .741}11 & \cellcolor[rgb]{ .635,  .635,  .635}15 & \cellcolor[rgb]{ .635,  .635,  .635}15 & \cellcolor[rgb]{ .675,  .675,  .675}13.5 \\
    $d_c$ & \cellcolor[rgb]{ .663,  .663,  .663}14 & \cellcolor[rgb]{ .686,  .686,  .686}13 & \cellcolor[rgb]{ .663,  .663,  .663}14 & \cellcolor[rgb]{ .686,  .686,  .686}13 & \cellcolor[rgb]{ .608,  .608,  .608}16 & \cellcolor[rgb]{ .714,  .714,  .714}12 & \cellcolor[rgb]{ .671,  .671,  .671}13.7 \\
    $K_u$ & \cellcolor[rgb]{ .584,  .584,  .584}17 & \cellcolor[rgb]{ .663,  .663,  .663}14 & \cellcolor[rgb]{ .714,  .714,  .714}12 & \cellcolor[rgb]{ .608,  .608,  .608}16 & \cellcolor[rgb]{ .714,  .714,  .714}12 & \cellcolor[rgb]{ .686,  .686,  .686}13 & \cellcolor[rgb]{ .663,  .663,  .663}14.0 \\
    $R_{s,d}$ & \cellcolor[rgb]{ .741,  .741,  .741}11 & \cellcolor[rgb]{ .584,  .584,  .584}17 & \cellcolor[rgb]{ .608,  .608,  .608}16 & \cellcolor[rgb]{ .635,  .635,  .635}15 & \cellcolor[rgb]{ .584,  .584,  .584}17 & \cellcolor[rgb]{ .557,  .557,  .557}18 & \cellcolor[rgb]{ .616,  .616,  .616}15.7 \\
    $K_c$ & \cellcolor[rgb]{ .557,  .557,  .557}18 & \cellcolor[rgb]{ .557,  .557,  .557}18 & \cellcolor[rgb]{ .584,  .584,  .584}17 & \cellcolor[rgb]{ .557,  .557,  .557}18 & \cellcolor[rgb]{ .557,  .557,  .557}18 & \cellcolor[rgb]{ .608,  .608,  .608}16 & \cellcolor[rgb]{ .569,  .569,  .569}17.5 \\
    $c_F$ & \cellcolor[rgb]{ .529,  .529,  .529}19 & \cellcolor[rgb]{ .529,  .529,  .529}19 & \cellcolor[rgb]{ .502,  .502,  .502}20 & \cellcolor[rgb]{ .529,  .529,  .529}19 & \cellcolor[rgb]{ .529,  .529,  .529}19 & \cellcolor[rgb]{ .529,  .529,  .529}19 & \cellcolor[rgb]{ .525,  .525,  .525}19.2 \\
    $\beta$ & \cellcolor[rgb]{ .502,  .502,  .502}20 & \cellcolor[rgb]{ .502,  .502,  .502}20 & \cellcolor[rgb]{ .529,  .529,  .529}19 & \cellcolor[rgb]{ .502,  .502,  .502}20 & \cellcolor[rgb]{ .502,  .502,  .502}20 & \cellcolor[rgb]{ .502,  .502,  .502}20 & \cellcolor[rgb]{ .51,  .51,  .51}19.8 \\
    \bottomrule
    \end{tabular}}%
  \label{tab:sens_rank}%
\end{table}%

Table~\ref{tab:sens_rank} highlights several out of trend parameter-simulation combinations. Parameters $d_F$ and $K_s$, which characterizes the slope of the lung recruitment function and the lung resistance, are at their most sensitive under normal breathing but drop in sensitivity under interventions. Conversely, $c_c$ is ranked lower sensitivity without CPAP but increases ranking noticeably with CPAP, and inductance $I$ appears to be at its most sensitive during increased $R_u$ conditions,  Finally, $R_{s,d}$ shows medium sensitivity under normal breathing and high $C_w$ but low sensitivity for all other simulations. These findings seem to indicate that the static respiratory compliance curves exhibit a larger influence over the breathing output during normal breathing, but influence shifts to the airway parameterization during increased $R_u$ during expiration. These features would be masked if only the overall ranking was used to determine sensitivity, though it still remains to discuss if these low-to-mid sensitivity shifts among simulations are sensitive enough to manifest in the outputs.

Parameters that exhibit low to medium sensitivity across all simulations include $C_{ve}$, $R_{ve}$, and $I$. We expect that while it would not be effective to remove lung tissue viscoelasticity and airway inertial effects from the model, the actual values of these parameters do not appear to affect the model outputs traditionally seen in experimental data and therefore could remain fixed at nominal values during an optimization.

Table~\ref{tab:subset} shows the subsets chosen for each of the six simulations as described previously in Section~\ref{sec:SS}, sorted in the same order as Table~\ref{tab:sens_rank} with a line separating the top six sensitive parameters by ranking. We search for a subset of parameters that is independent for all or most of the six simulations, with parameter values that can distinguish between simulations. It is initially clear that $\gamma$, $k$, $R_{u,mult}$, $A_{mus}$, and $d_w$ were chosen for all simulations and should be considered a candidate for an independent subset. These describe the lung and chest wall compliance curves, the amplitude of the respiratory muscle driving pressure, and the level of airway braking. $K_u$ is also chosen for all simulations. However, it is considered to have low sensitivity, and therefore attempts to optimize it would be both unnecessary and potentially hinder the computation. We also note that $R_{u,m}$ is chosen for all simulations except high $C_w$ with no interventions. Because it is also a highly sensitive parameter, we consider it as a candidate for the optimized subset. It is also interesting to see the increase in number of identifiable parameters with the addition of increased $R_u$ on expiration, possibly because the shape of the output differs from the other two interventions. 

Considering all of the above, the final independent sensitive candidate parameter set was estimated using the Levenberg-Marquardt optimization algorithm as described in~\ref{sec:opt}:
\begin{equation}
\mu_{est}=\{\gamma, k,R_{u,m}, R_{u,mult},A_{mus},d_w\}
\end{equation}

\begin{table}[htbp]
 \caption{Display of parameters chosen by subset selection for each simulation. Parameters are ordered identical to Table~\ref{tab:sens_rank} for comparison. Note that $K_u$ was chosen for all six simulations but is of low sensitivity. }
	{\footnotesize
    \begin{tabular}{l|ccc|ccc|c}
\toprule
          & \multicolumn{3}{c|}{High $C_w$} & \multicolumn{3}{c|}{Low $C_w$} & Number \\
    \multicolumn{1}{c|}{Parameter} & normal $R_u$ & increased $R_u$ & $P_{ao}=5$ & normal $R_u$ & increased $R_u$ & $P_{ao}=5$ & chosen \\
    \midrule
    $\gamma$ & X     & X     & X     & X     & X     & X     & 6 \\
    $k$   & X     & X     & X     & X     & X     & X     & 6 \\
    $R_{u,m}$ &       & X     & X     & X     & X     & X     & 5 \\
    $R_{u,mult}$ & X     & X     & X     & X     & X     & X     & 6 \\
    $A_{mus}$ & X     & X     & X     & X     & X     & X     & 6 \\
    $d_w$ & X     & X     & X     & X     & X     & X     & 6 \\
    \midrule
    $R_{s,m}$ &       & X     & X     &       & X     &       & 3 \\
    $C_{ve}$ &       & X     &       &       & X     &       & 2 \\
    $V_{c,max}$ & X     &       &       &       &       &       & 1 \\
    $R_{ve}$ & X     & X     &       & X     & X     &       & 4 \\
    $d_F$ & X     & X     &       & X     & X     &       & 4 \\
    $K_s$ & X     & X     &       & X     & X     & X     & 5 \\
    $I$   &       & X     &       &       & X     &       & 2 \\
    $c_c$ &       & X     & X     & X     & X     & X     & 5 \\
    $d_c$ &       & X     & X     &       & X     & X     & 4 \\
    $K_u$ & X     & X     & X     & X     & X     & X     & 6 \\
    $R_{s,d}$ & X     & X     & X     & X     & X     &       & 5 \\
    $K_c$ &       &       &       &       &       &       & 0 \\
    $c_F$ &       &       &       &       &       &       & 0 \\
    $\beta$ &       &       &       &       &       &       & 0 \\
    \bottomrule
    \end{tabular}}%
  \label{tab:subset}%
\end{table}%


\subsection{Optimizations}

Each of the optimizations for all three levels of added noise converged according to the Levenberg-Marquardt termination criteria. Figures~\ref{fig:gradnorm_G2},~\ref{fig:gradnorm_G5}, and~\ref{fig:gradnorm_G10} in Appendix A depict the decrease of the gradient norm with increasing iterations for the 100 optimizations performed on each simulation, where is seen that the gradient norm is near zero within 40 iterations. It also appears that the two simulations with increased expired $R_u$ (middle row) had a greater fraction of optimizations that took more iterations, compared to the other four simulations where the majority converged in under 10 iterations. Several optimizations showing a transient increase in gradient norm before finally approaching 0, e.g. high $C_w$ with no interventions. 

\begin{figure}[h]
\includegraphics[scale=0.8]{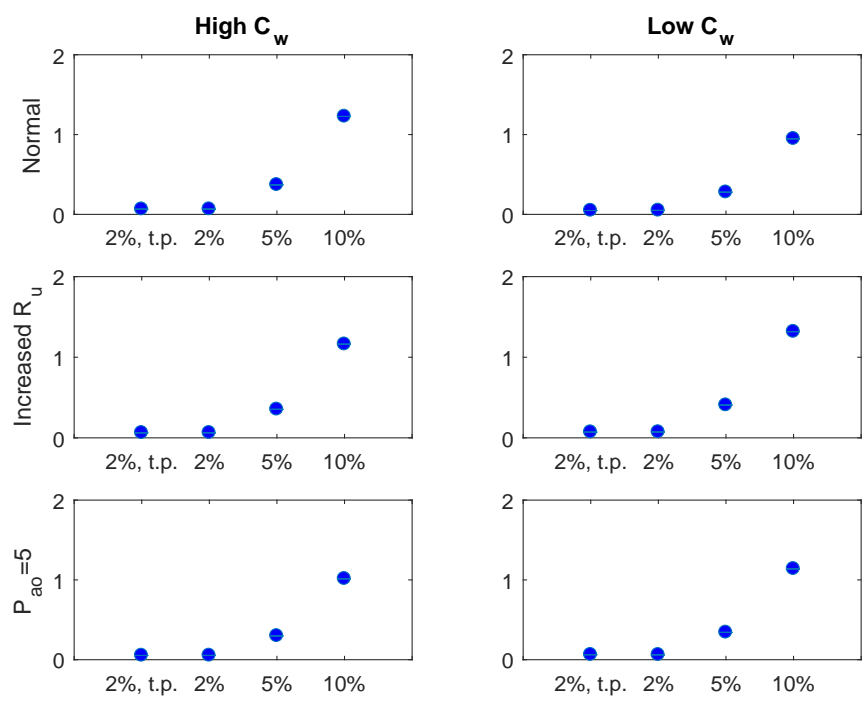}
\caption{Average final cost obtained from 100 optimizations at each of three levels of noise-added pseudo-data. The first data point in each is the mean cost using the true parameters (t.p.). } 
\label{fig:costave_noise}
\end{figure}

\begin{figure}[h]
\includegraphics[scale=0.75]{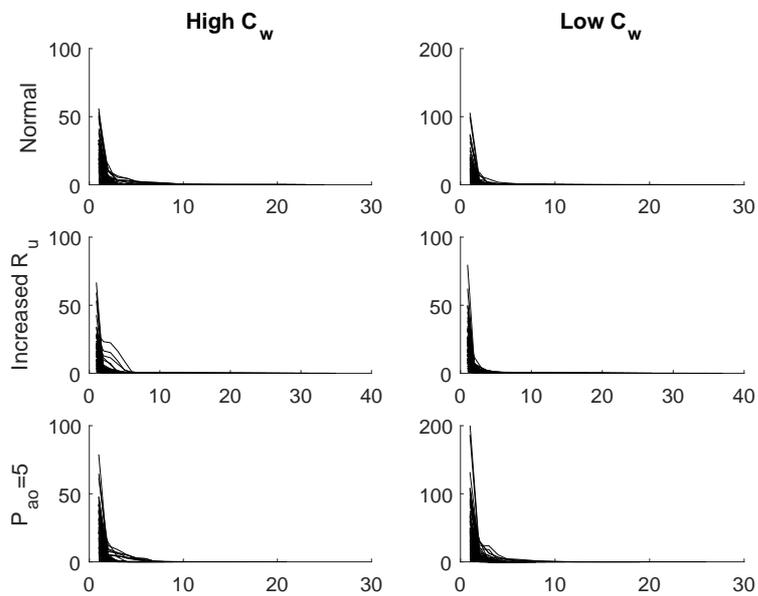}
\caption{Cost at each iteration for 100 L-M optimizations done on pseudo-data with 2\% noise. } 
\label{fig:cost_G2}
\end{figure}

\begin{figure}[h]
\includegraphics[scale=0.75]{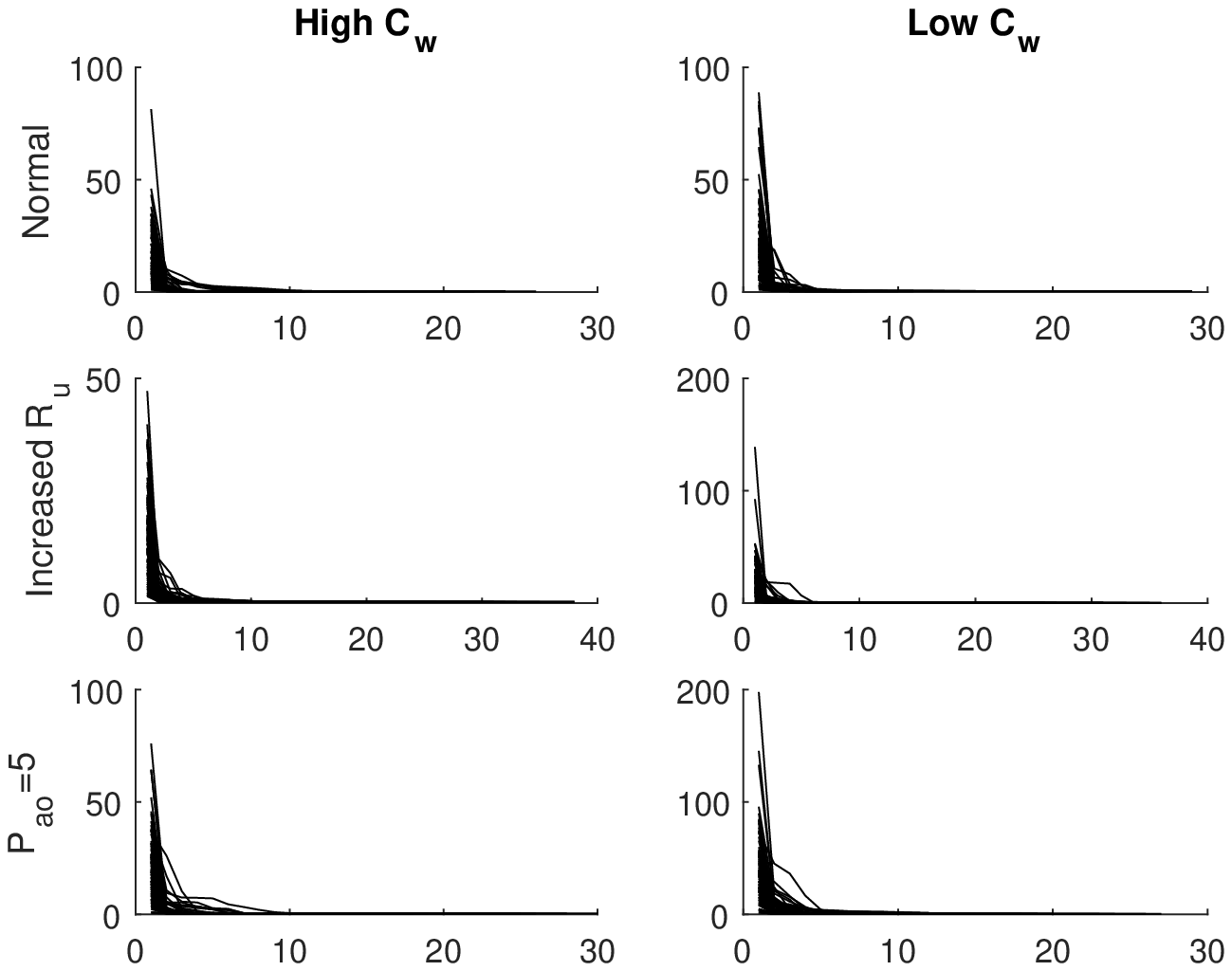}
\caption{Cost at each iteration for 100 L-M optimizations done on pseudo-data with 5\% Gaussian noise. } 
\label{fig:cost_G5}
\end{figure}

\begin{figure}[h]
\includegraphics[scale=0.75]{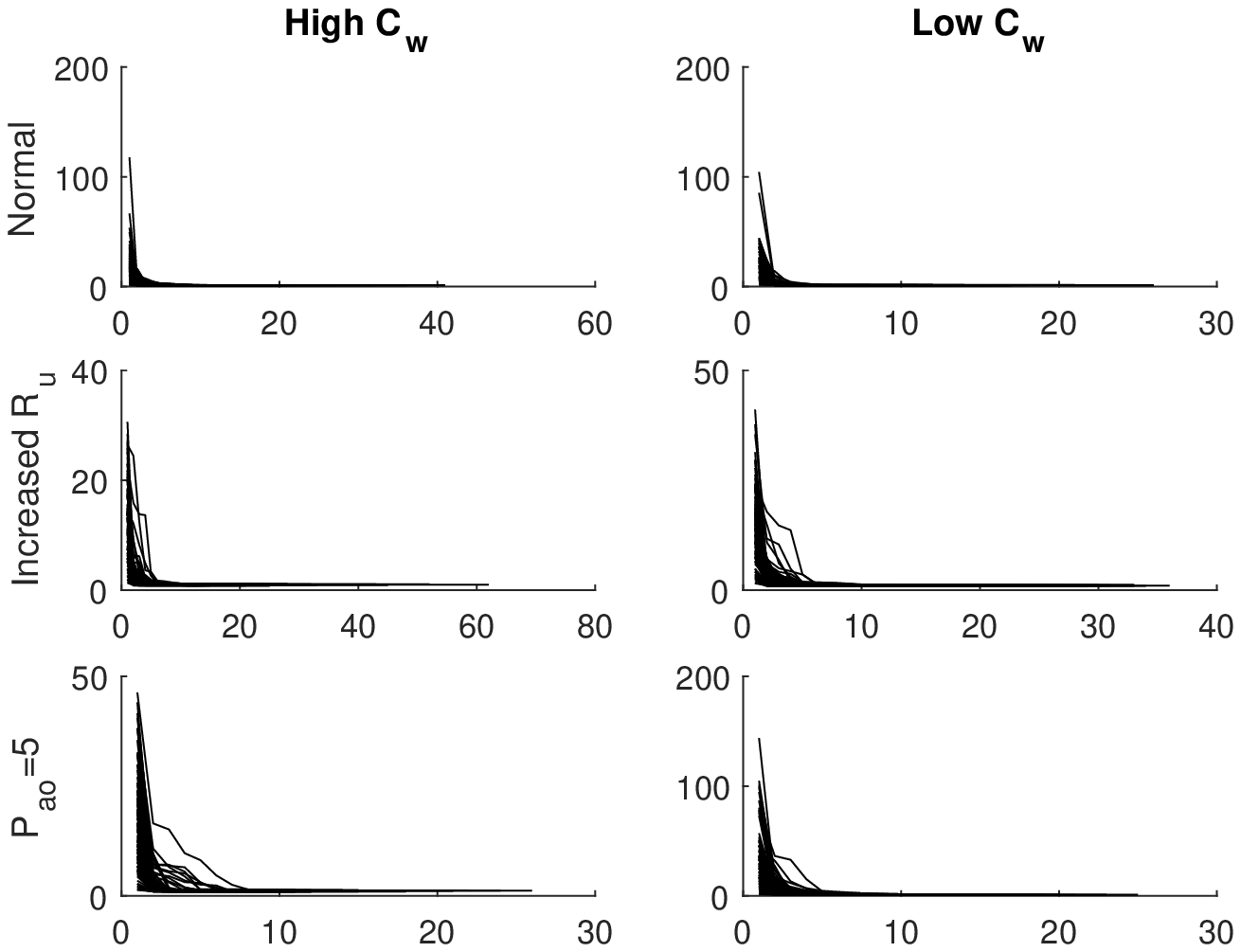}
\caption{Cost at each iteration for 100 L-M optimizations done on pseudo-data with 10\% Gaussian noise. } 
\label{fig:cost_G10}
\end{figure}

The mean final cost for each set of 100 realization of pseudo-data is shown in Figure~\ref{fig:costave_noise}. To a check that comparable cost would be attained using true parameters, 50 additional optimizations for each simulation condition using 2\% noise-added pseudo-data were run. Results indicate that the cost decreased to reasonable minumum in each optimization. For completeness, the cost decrease is also shown in Figures~\ref{fig:cost_G2},~\ref{fig:cost_G5}, and~\ref{fig:cost_G10}. 

\begin{landscape}
\begin{figure}
\includegraphics[scale=0.5]{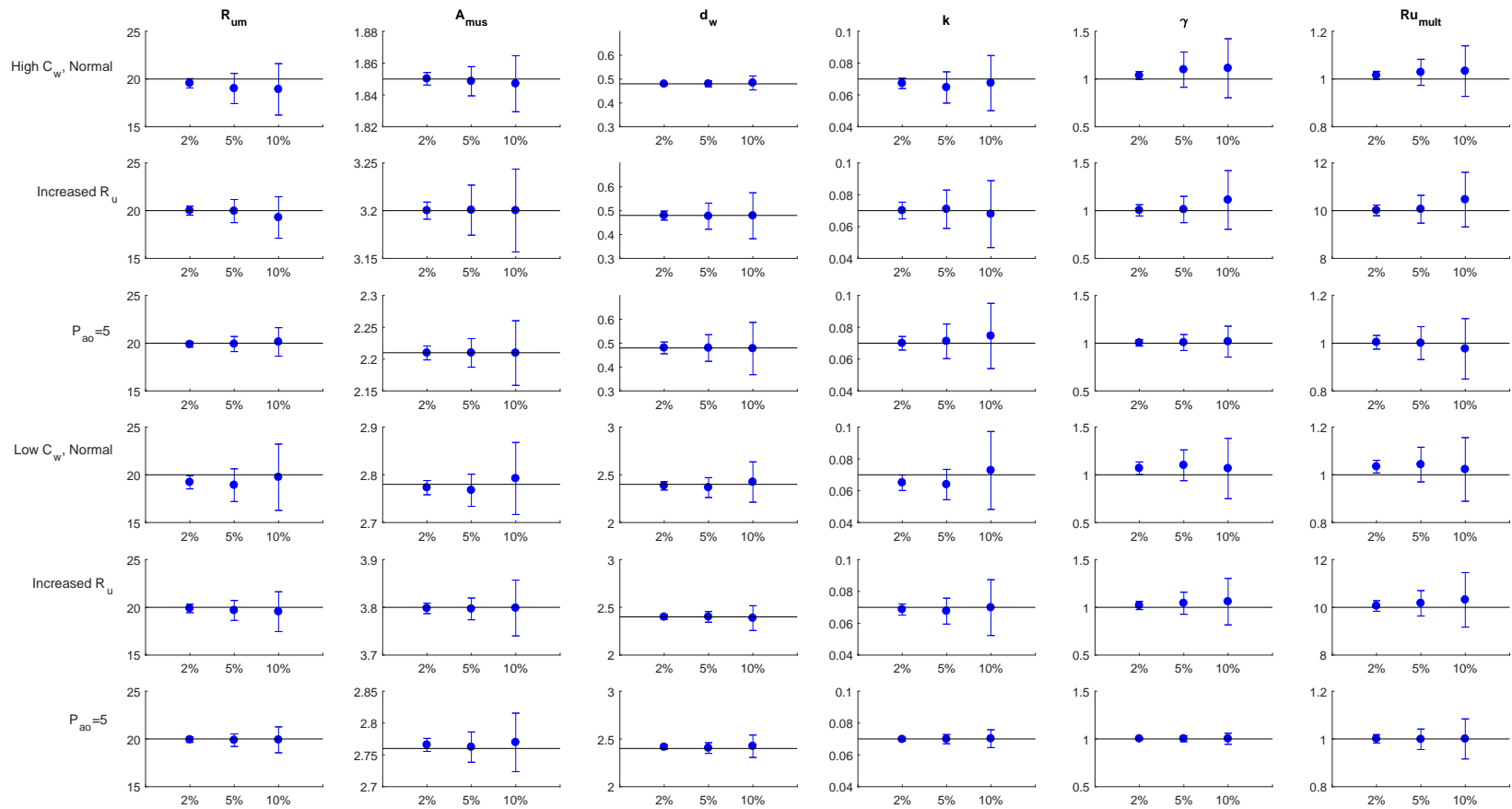}
\caption{Mean parameter values obtained from 100 optimizations at each of three levels of noise-added pseudo-data. Horizontal black lines in each panel indicate the true parameter value. Error bars represent 1 standard deviation.} 
\label{fig:all_means}
\end{figure}
\end{landscape}

\begin{landscape}
\begin{figure}[h]
\includegraphics[scale=0.5]{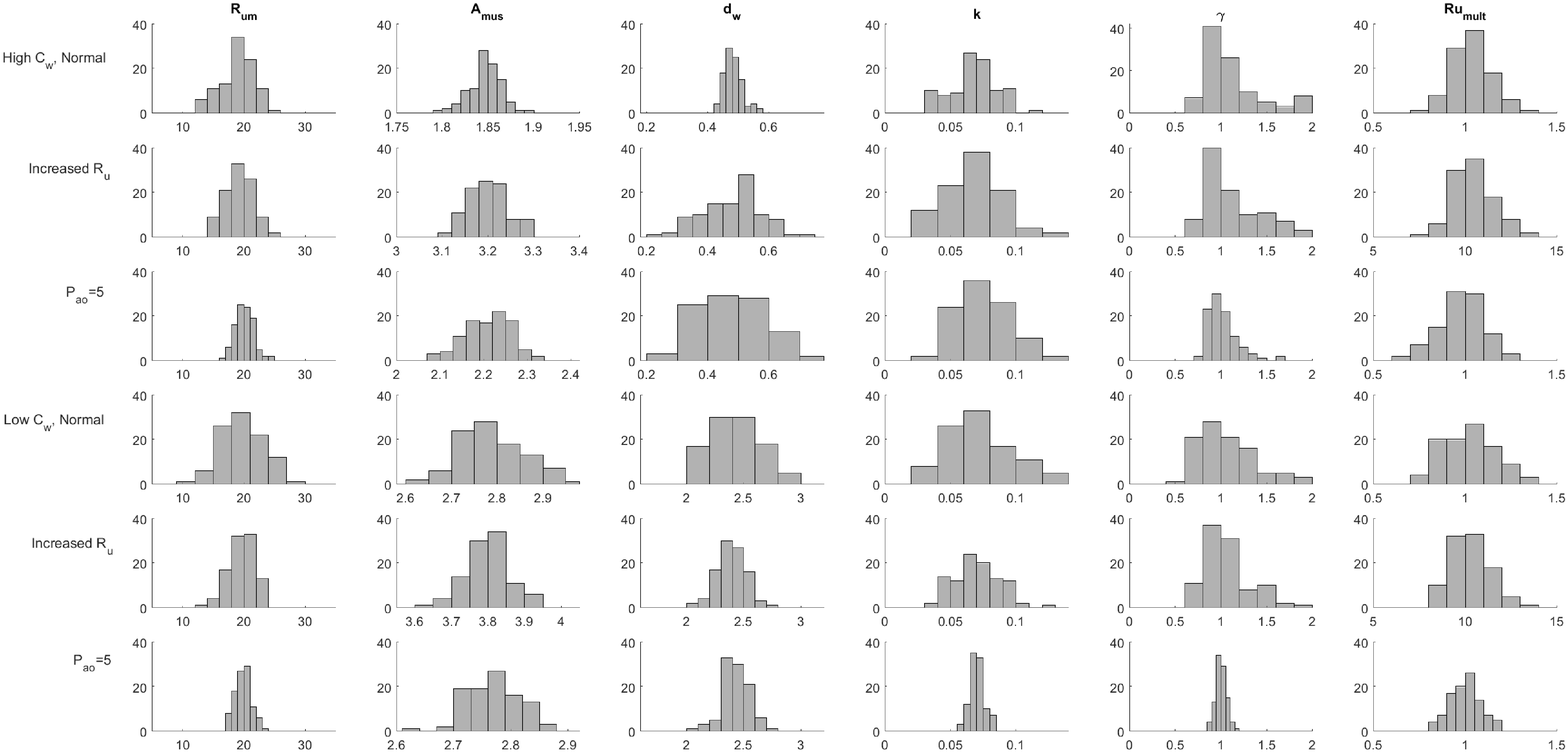}
\caption{Histograms for each parameter and simulation from 100 optimizations done on pseudo-data with 10\% noise.} 
\label{fig:hist_G10}
\end{figure}
\end{landscape}

Table~\ref{tab:simulations} gives the mean optimized parameter values and standard deviations for each simulation, as also depicted in Fig.~\ref{fig:all_means}.  Note that true parameter values for $R_{u,m}$, $k$, and $\gamma$ are the same for all simulation conditions but $A_{mus}$, $d_w$, and $R_{u,mult}$ differ depending on the simulation conditions. On average, optimizations performed on pseudo-data with 2\% noise were able to reasonable identify values for the six parameters within $\sim$8\%. Parameter value estimates became less precise as the noise level increased. The spread of parameter estimate values is also depicted in the histograms in Appendix B, Figures~\ref{fig:hist_G2},~\ref{fig:hist_G5}, and~\ref{fig:hist_G10}. 

\begin{figure}[h]
\includegraphics[scale=0.8]{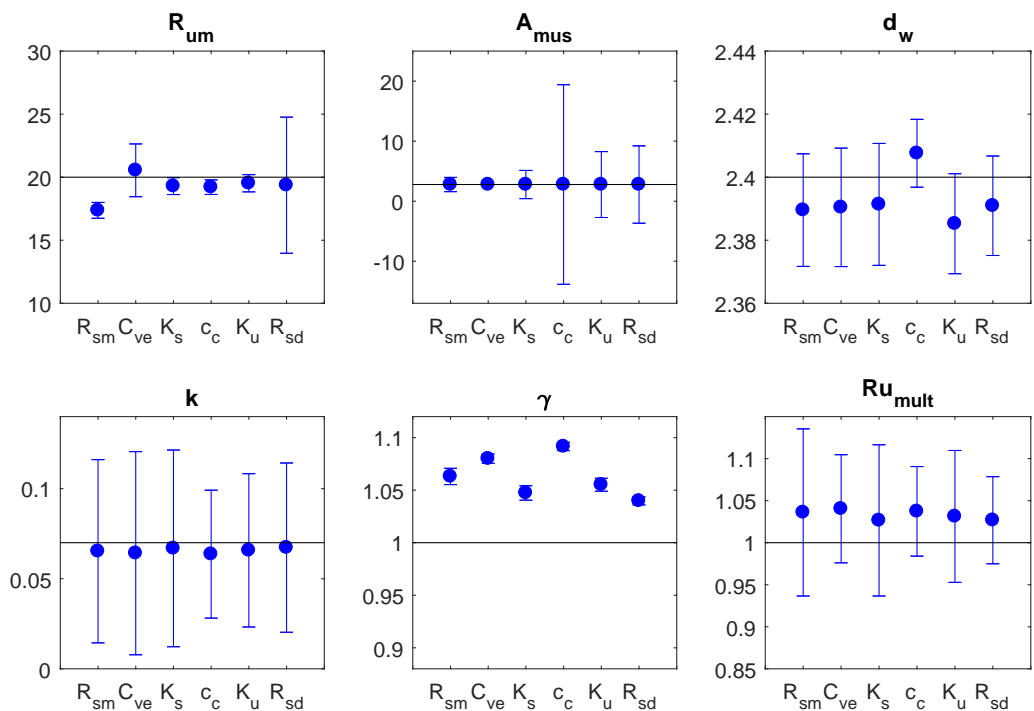}
\caption{Means and standard deviations of the original six parameters in $\mu_{est}$ when a 7th parameter was added, indicated on the x-axis. Simulation condition was low $C_w$, normal $R_u$, and zero CPAP, using 2\% noise-added pseudo-data. Horizontal black lines in each panel indicate the true parameter value. Error bars represent 1 standard deviation. } 
\label{fig:means_7params}
\end{figure}

The effect of adding an additional parameter on parameter estimate precision that is not in the independent sensitive subset $\mu_{est}$  was explored by adding sensitive and/or non-identifiable parameters $R_{sm},C_{ve},K_s,c_c,K_u$, and $R_{sd}$ individually as the 7th parameter included in the optimization. The first two are the 7th and 8th most sensitive parameters, and the last four are additional parameters that were chosen in the identifiable subset for 5 or 6 simulations but were not included in $\mu_{est}$ because of the lack of sensitivity. This analysis was done on the simulation condition of low $C_w$, normal $R_u$, and zero CPAP, using 2\% noise-added pseudo-data as an illustration.  Figure~\ref{fig:means_7params} shows the means and standard deviations of the original six in $\mu_{est}$ when the 7th parameter was added, indicated on the x-axis of each panel. The effect of the 7th parameter has varying effects, for example the standard deviation of $\gamma$ remains small but the mean estimate is consistently high. In contrast, estimates for $k$ hover around the mean of 0.07 but are remarkably less precise for all 7th parameters than with just optimizatin the 6 parameters. 

We highlight evidence of some possible correlations that may still be present.  Under low $C_w$ and no intervention, mean $R_{u,m}$ and $k$ were at their lowest values of the six simulation conditions while $\gamma$ and $R_{u,mult}$ were at their highest. Since $k$ and $\gamma$ describe different parts of the overall lung compliance curve, and $R_{u,m}$ and $R_{u,mult}$ both affect model output at different portions of the breathing cycle, it is understandable that their values may be loosely correlated. Despite this, the optimizer still converged every time and attained values close to the nominal parameters that generated the original data.

\begin{table}[!ht]
  \caption{Mean parameter values from 100 optimization runs for each of the six simulations, reported with standard deviation (SD).}
		{\footnotesize
    \begin{tabular}{cc|cccccc}
\hline\noalign{\smallskip}
										&																					&				&			\multicolumn{4}{c}{Mean optimized parameter values (SD), 2\% gaussian noise}  & \\
    {\bf $C_w$} 	& {\textbf{Intervention}} & $R_{um}$ 		& $A_{mus}$ 		& $d_w$ 			& $k$ 					& $\gamma$ 		& $R_{u,mult}$ \\
\noalign{\smallskip}
\hline\noalign{\smallskip}
    \multirow{3}{*}{High}						& none 										& 	19.6 (.5)	& 1.850 (.004)	& 0.479 (.005) & 0.067 (.003)	& 1.04 (.04)	& 1.02 (.02)\\
																		& Incr exp $R_u$ 					&	 	20.0 (.5) & 3.200 (.009)	& 0.479	(.019) & 0.070	(.005) & 1.00 (.06)	& 10.0 (.2)\\
																		& $P_{ao}=5$							&	 	19.9 (.3) & 2.210 (.011) 	& 0.480	(.025) & 0.070	(.004) & 1.00 (.03)	& 1.00 (.03)\\
    \midrule
    \multirow{3}{*}{Low } 					& none 										&  	19.2 (.7) & 2.773 (.015)	& 2.39	(.04) & 0.065	(.005) & 1.07 (.07) & 1.03 (.03) \\
																		& Incr exp $R_u$ 					&	 	19.9 (.4) & 3.800 (.011)	& 2.40	(.02) & 0.069	(.004) & 1.02 (.04) & 10.1 (.2)\\
																		& $P_{ao}=5$							&	 	19.9 (.3) & 2.766 (.010)	& 2.41	(.02) & 0.070	(.001) & 1.00	(.01) & 1.00 (.02)\\
\noalign{\smallskip}
\hline\noalign{\smallskip}
										&																					&				&			\multicolumn{4}{c}{Mean optimized parameter values (SD), 5\% gaussian noise}  & \\
    {\bf $C_w$} 	& {\textbf{Intervention}} & $R_{um}$ 		& $A_{mus}$ 		& $d_w$ 			& $k$ 					& $\gamma$ 		& $R_{u,mult}$ \\
\noalign{\smallskip}\hline\noalign{\smallskip}
    \multirow{3}{*}{High}						& none 										& 	19.0 (1.6)	& 1.849 (.009)& 0.480 (.013) & 0.065 (.010)	& 1.10 (.18)	& 1.03(.05)\\
																		& Incr exp $R_u$ 					&	 	20.0 (1.2) & 3.200 (.026)	& 0.477	(.055) & 0.071	(.012) & 1.01 (.14)	& 10.1 (.6)\\
																		& $P_{ao}=5$							&	 	19.9 (.8) & 2.210 (.022) 	& 0.480	(.056) & 0.071	(.011) & 1.01 (.08)	& 1.00 (.07)\\
    \midrule
    \multirow{3}{*}{Low } 					& none 										&  	18.9 (1.7) & 2.768 (.034)	& 2.36	(.10) & 0.064	(.010) & 1.10 (.16) & 1.04 (.07) \\
																		& Incr exp $R_u$ 					&	 	19.7 (1.0) & 3.797 (.023)	& 2.40	(.06) & 0.068	(.008) & 1.04 (.11) & 10.2 (.5)\\
																		& $P_{ao}=5$							&	 	19.9 (.7) & 2.762 (.024)	& 2.40	(.06) & 0.070	(.003) & 1.00	(.03) & 1.00 (.04)\\
\noalign{\smallskip}
\hline\noalign{\smallskip}
										&																					&				&			\multicolumn{4}{c}{Mean optimized parameter values (SD), 10\% gaussian noise}  & \\
    {\bf $C_w$} 	& {\textbf{Intervention}} & $R_{um}$ 		& $A_{mus}$ 		& $d_w$ 			& $k$ 					& $\gamma$ 		& $R_{u,mult}$ \\
\noalign{\smallskip}\hline\noalign{\smallskip}
    \multirow{3}{*}{High}						& none 										& 	18.9 (2.7)	& 1.847 (.018)& 0.484 (.029) & 0.067 (.017)	& 1.11 (.31)	& 1.03 (.11)\\
																		& Incr exp $R_u$ 					&	 	19.3 (2.2) & 3.200 (.043)	& 0.479	(.096) & 0.068	(.021) & 1.11 (.31)	& 10.5 (1.1)\\
																		& $P_{ao}=5$							&	 	20.1 (1.5) & 2.210 (.051)	& 0.478	(.110) & 0.075	(.021) & 1.02 (.16)	& 0.98 (.13)\\
    \midrule
    \multirow{3}{*}{Low } 					& none 										&  	19.7 (3.5) & 2.792 (.076)	& 2.42	(.21) & 0.073	(.025) & 1.07 (.32) & 1.02 (.13) \\
																		& Incr exp $R_u$ 					&	 	19.5 (2.1) & 3.799 (.059)	& 2.39	(.13) & 0.070	(.018) & 1.06 (.24) & 10.3 (1.1)\\
																		& $P_{ao}=5$							&	 	19.9 (1.4) & 2.770 (.046)	& 2.42	(.12) & 0.070	(.006) & 1.00	(.063) & 1.00 (.08)\\
\noalign{\smallskip}\hline
    \end{tabular}}
  \label{tab:simulations}
\end{table}


\section{Discussion}\label{sec:discuss}

This study showed that a combination of sensitivity analysis and subset selection can identify an independent sensitive subset of six parameters characterizing a respiratory mechanics model under six simulation conditions. Pseudo-data generated from simulated outputs of airflow and pleural pressure were used with perturbed nominal parameter values to test the ability of a gradient-based optimization algorithm to estimate parameters close to nominal values. Nominal parameter values were generated in the context of an idealized preterm infant. Parameters and simulation conditions that are quantifiable ahead of time, such as estimates of static lung volumes based on subject size or amount of ventilation assistance, were kept fixed during optimizations. The parameters associated with amplitude of breathing ($A_{mus}$), level of grunting ($R_{u,mult}$) and degree of chest wall compliance ($d_w$) that differentiated between simulations as well as 17 parameters that were set the same for all simulations were analyzed computationally . 

\begin{landscape}
\begin{figure}[h]
\includegraphics[scale=0.5]{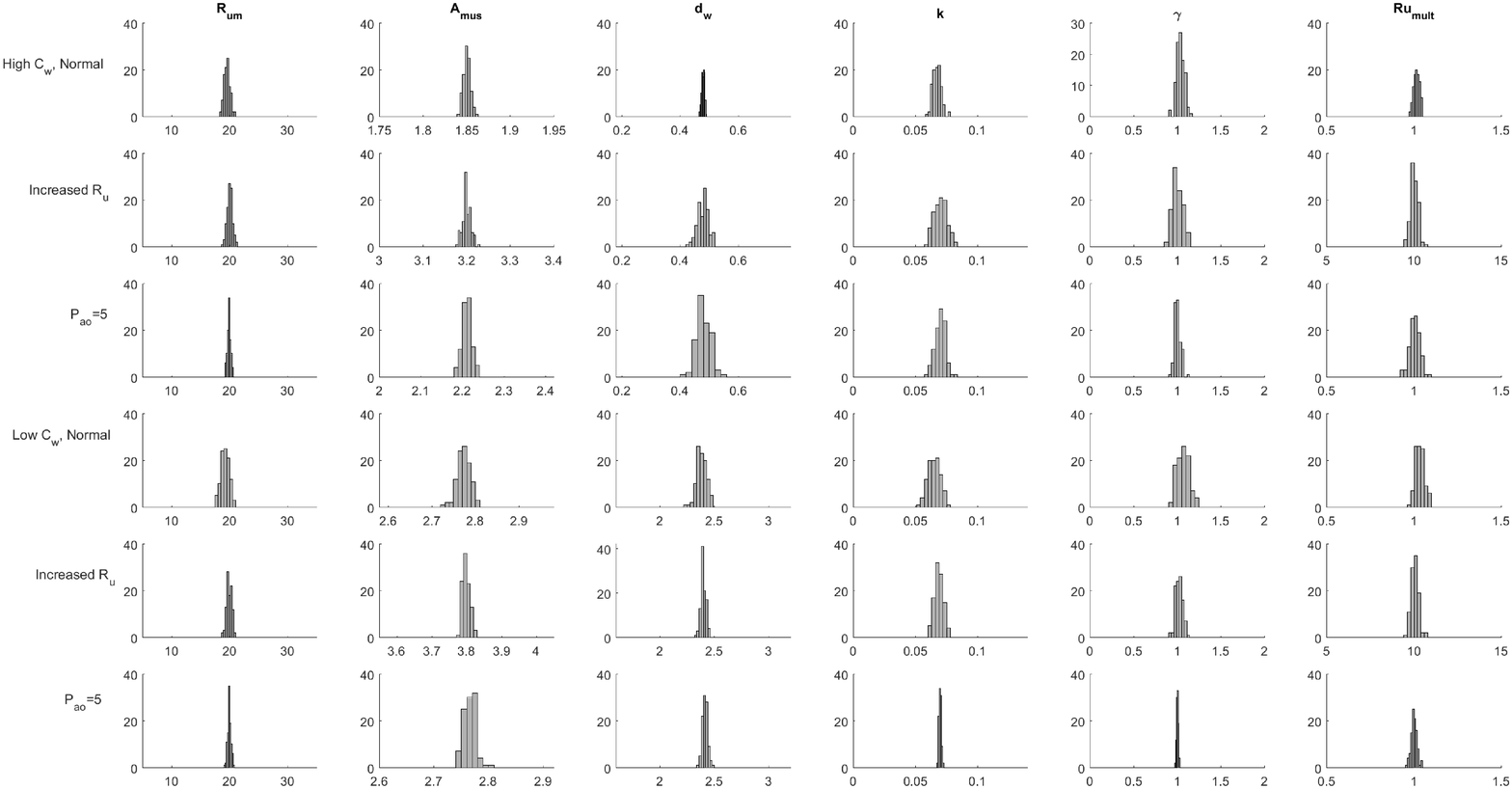}
\caption{Histograms for each parameter and simulation from 100 optimizations done on pseudo-data with 2\% noise.} 
\label{fig:hist_G2}
\end{figure}
\end{landscape}

Of the six most sensitive parameters as identified by composite relative sensitivities, all were chosen by subset selection as independent for all six simulations except for $R_{u,m}$, which was not chosen for the high $C_w$ normal $R_{u}$ condition (see Table~\ref{tab:subset} column 1). Though it was not chosen, the mean reported for $R_{u,m}$ in Table~\ref{tab:simulations} is close to the nominal value 20 with a reasonable standard deviation similar to the other five simulations, also supported by the histogram for $R_{u,m}$ in row 1 of Fig.~\ref{fig:hist_G2}. Because $K_u$ was chosen as independent for all six simulations we initially included it in optimizations (results not shown). However, since the output vector $y$ is not sensitive to $K_u$, its value is not critical to parameter estimation and indeed the the optimized value of $K_u$ fluctuated greatly. Further evidence of the impact of $K_u$ and the other additional parameters on the optimization of six parameters in $\mu_{est}$ was given in Fig~\ref{fig:means_7params}. The final parameter set $\gamma,k,R_{u,m},R_{u,mult},A_{mus},d_w$ characterizes the static respiratory compliance curves that underlie the dynamics, airway resistance, and respiratory muscle pressure amplitude. 

Knowledge about the parameters considered to be less sensitive or badly identifiable could be used to simplify the model. As an example, the sigmoidal equation for fractional recruitment includes parameters $\gamma$, $d_F$, $c_F$, and $\beta$, written in order of decreasing sensitivity. Insensitive parameters $\beta$ and $c_F$ were also not chosen as independent for any of the six simulations, leading to the conclusion that fractional recruitment is fully defined by $\gamma$, the maximum fractional recruitment, and $d_F$, a scalar characterizing the slope at maximum recruitment rate (see Tables~\ref{tab:glossary} and~\ref{tab:functions}). Note also that $d_F$ was not chosen in the identifiable subset in the two simulations with CPAP ($P_{ao}=5$). This is likely because the higher airway pressure induces higher lung elastic recoil $P_{el}$ (see Fig.~\ref{fig:states}), putting the operating point of the recruitment function $F_{rec}$ closer to the upper tail of the sigmoid where its slope $d_F$ matters less. A second example is the the parabolic equation for collapsible airways resistance $R_c(V_c)$, defined by parameters $V_{c,max}$ and $K_c$ which were chosen in the independent identifiable subset in 1 and 0 simulations respectively (see Table~\ref{tab:subset}). This is unsurprising given that the parabolic effects of $V_c$ on $R_c$ are most impactful when breathing is at the extremes, but infant breathing is generally maintained at tidal volumes where $R_c$ is likely linear or even constant. However, if experimental data was acquired during a condition exhibiting situation expiratory flow limitation such as obstructive pulmonary disease~\citep{Khirani01}, $R_c$ may play a bigger role. These avenues of model simplification can be explored in future investigations.

This study focuses on local differential sensitivity analysis, SVD/QR-based subset selection, and gradient-based optimization. Nominal values were determined based on a priori knowledge of the physiological system to simulate airflow and pleural pressure typical of a surfactant-treated 1kg infant, as was published previously~\cite{Ellwein18}. Performing the sensitivity analysis with six different but common simulation conditions, for which a subset of parameters vary greatly, begins to explore the issues surrounding the sensitivity of model output to vastly different parameter values in the viable parameter space for the system. There are numerous approaches available to study parameter identification and estimation with optimization that were not used in this study. A global sensitivity analysis, such as performing a repeated local analysis using random uniform or Latin hypercube sampling~\cite{Raue13,Olsen18} or a using a global method like Sobol indices or Morris elementary effects~\cite{LeRolle13,Olsen18}, would explore a greater extent of the parameter space which may allow for more rigorous future investigation of real data obtained under a variety of conditions. Olufsen and Ottesen~\cite{Olufsen13} compare parameter identifiability of a model of heart rate regulation using a structured correlation method, the SVD/QR method, and model Hessian subspace method. Their work found the structured correlation method to produce the ``best'' subset with fewest interdependent parameters, and the SVD/QR method did not give as precise parameter estimates but it much more computationally feasible.  

It should be noted that the fixed set of parameters ${TLC,RV,VC,FRC,RR,f,T,V_0,\alpha,\nu\,c_w}$ are not explored with either the sensitivity analysis or subset selection. Some of these parameters would be estimated a priori for any patient using this model, such as the static lung volumes calculated from patient anthropometric measurements. However, parameters such as $V_0$ that characterize a constitutive relationship in the model are estimated based on literature averages and could vary between patients, but were kept constant because of parameter dependencies. Examining the effects of analyses on fixed parameters is an avenue for future investigations especially when clinical data becomes available.

 Several features of preterm infant respiratory mechanics are not yet captured by the model and will be addressed in future modifications. These include variable frequency of breathing, non-sinusoidal respiratory muscle pressure, intermittent deep breathing (sighing), variable time spent in inspiration vs. expiration, paradoxical chest movement, and chemoreflex feedback. The addition of model modifications likely increases the parameter space, making it more critical to address the question of parameter identifiability. We will explore using these results to simplify constitutive relationships whose nonlinearity may not manifest during quiet tidal breathing typical of a newborn infant.

The clinical applicability of these analyses is directly related to both the available data and the model construction. As stated earlier, lung volume $V_A$ is often a piece of clinically available data, but is recorded as volume relative to FRC instead of an absolute lung volume. Any breath-to-breath changes in FRC are not captured in this data. Though the dynamic absolute $V_A$ could not be compared against the model output as a result, tidal volume (volume inspired in a single breath) could be added to the output vector for parameter estimation against clinical data. It is important to reiterate that we estimated parameters responsible during steady periodic breathing prior to any volume loss. If data was acquired during progressive volume loss, the results would likely change. It is evident from the subset selection results from increased expired $R_u$ that a slight change in the qualitative nature of the outputs increases the number of independent parameters. A future translational approach could follow what has been done by Kretschmer et al~\cite{Kretschmer17}, who compared model-based parameter estimation with a standard clinical method to determine compliances and resistances under several respiratory manuevers. From a different perspective, if a parameter determined to be non-identifiable was also deemed clinically relevant, this could motivate collection of new experimental data.

Respiratory mechanics models have been investigated for several years and many formulations exist; the current goal is the customization at the patient-specific level via parameter estimation performed on dynamic data tracings. This becomes an even greater challenge when working with a fragile population such as extremely preterm infants for whom experimental data collection is limited. This study indicates the feasibility of parameter estimation under a variety of experimental conditions. These methods will be applied to future data obtained in the NICU to estimate patient-specific parameters that may help uncover factors leading to progressive volume loss. The ability to predict volume loss could lead to prevention strategies and assist in the health and stability of the preterm infant population.

\section*{Data Availability}
The data and computer code used to support the findings of this study are available from the author upon request.

\section*{Conflicts of Interest}
The author declares no conflicts of interest.

\section*{Acknowledgments}
This research was supported in part by the Atlantic Pediatric Device Consortium FDA grant 5P50FD004193-07.


\newpage

\section*{Appendix}

\subsection*{Appendix A: Iteration and cost histories}\label{appendixA}

\begin{figure}[h]
\includegraphics[scale=0.5]{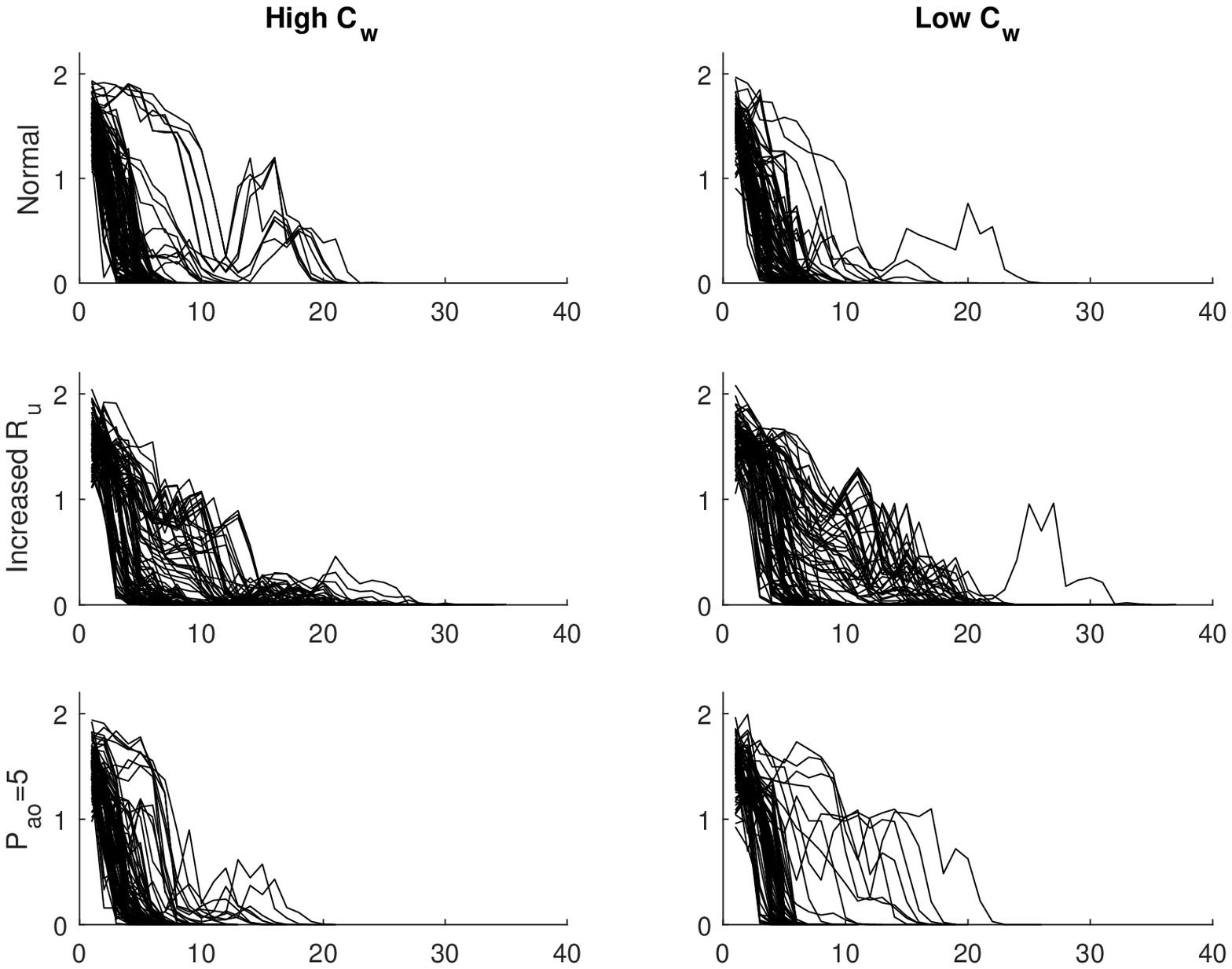}
\caption{Gradient norm at each iteration for 100 L-M optimizations done on pseudo-data with 2\% noise. } 
\label{fig:gradnorm_G2}
\end{figure}

\begin{figure}[h]
\includegraphics[scale=0.5]{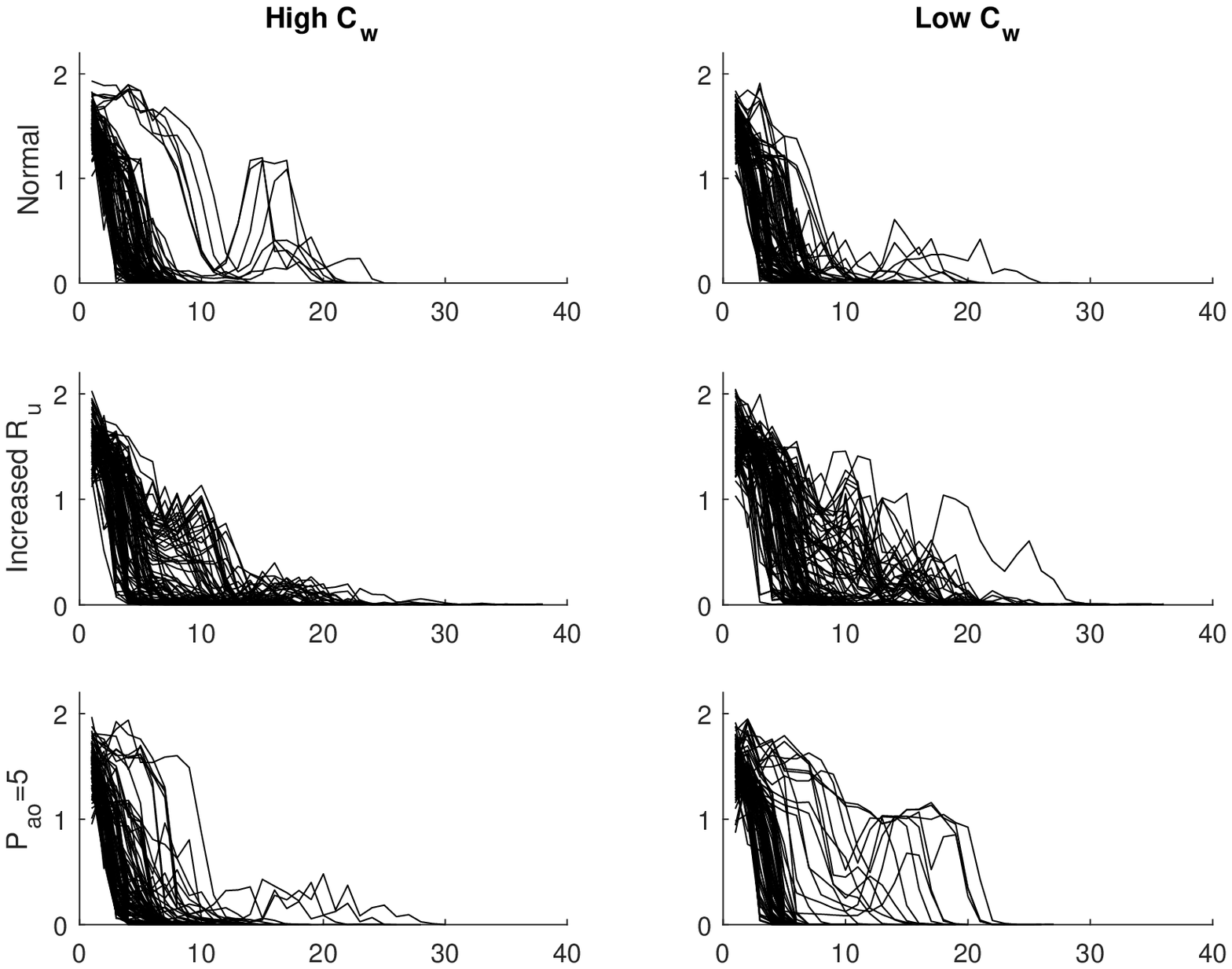}
\caption{Gradient norm at each iteration for 100 L-M optimizations done on pseudo-data with 5\% Gaussian noise. } 
\label{fig:gradnorm_G5}
\end{figure}

\begin{figure}
\includegraphics[scale=0.6]{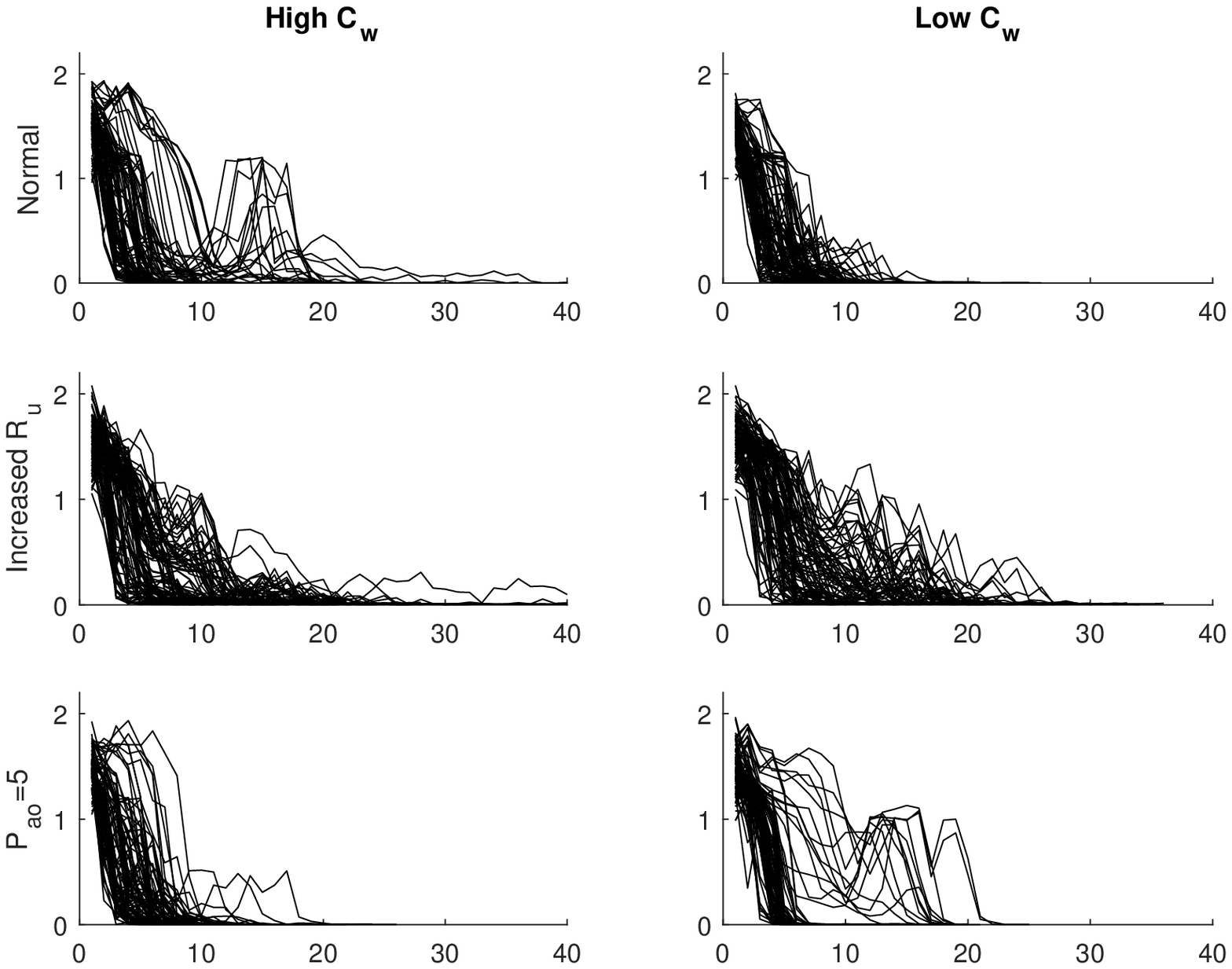}
\caption{Gradient norm at each iteration for 100 L-M optimizations done on pseudo-data with 10\% Gaussian noise. } 
\label{fig:gradnorm_G10}
\end{figure}

\clearpage
\newpage

\subsection*{Appendix B: Histograms}\label{appendixB}

\begin{landscape}
\begin{figure}[h]
\includegraphics[scale=0.5]{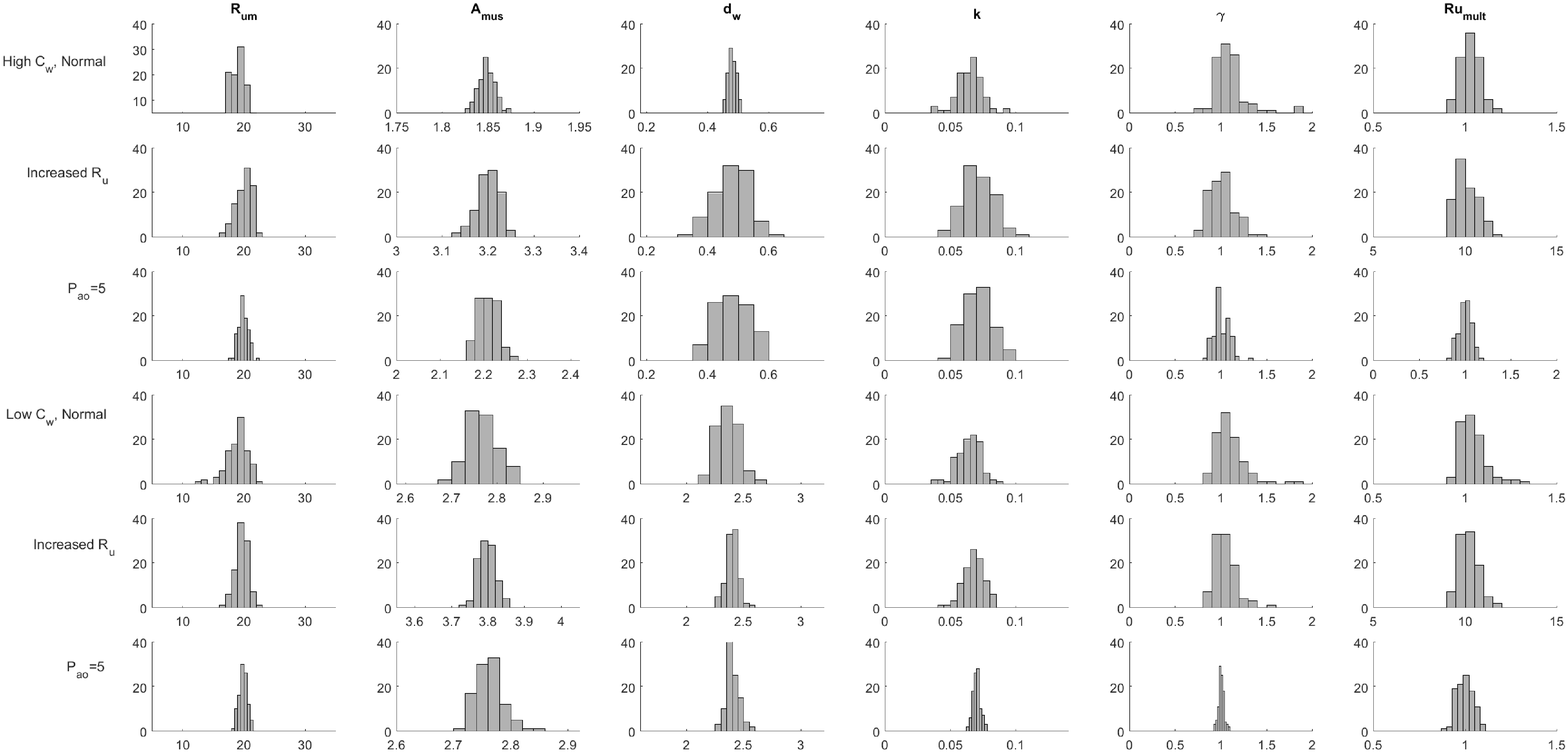}
\caption{Histograms for each parameter and simulation from 100 optimizations done on pseudo-data with 5\% noise.} 
\label{fig:hist_G5}
\end{figure}
\end{landscape}

\end{document}